\documentclass[useAMS,usenatbib]{mn2e}
\usepackage{natbib, graphicx, times}
\usepackage{amsmath}
\usepackage{mathtools}
\usepackage{array}
\usepackage{epstopdf}
\usepackage{amssymb}
\usepackage{extarrows}
\usepackage{color}
\usepackage{enumitem}
\usepackage{scrextend}
\usepackage{threeparttable}
\usepackage{scalerel}

\setlist{leftmargin=7.5mm}

\newcommand{\lsim}{\mathrel{\rlap{\lower4pt\hbox{\hskip1pt$\sim$}}
    \raise1pt\hbox{$<$}}}                
\newcommand{\gsim}{\mathrel{\rlap{\lower4pt\hbox{\hskip1pt$\sim$}}
    \raise1pt\hbox{$>$}}}                

\def\vecsign#1{\rule[1.388\LMex]{\dimexpr#1-2.5pt}{.36\LMpt}%
  \kern-6.0\LMpt\mathchar"017E}

\newcommand{\cbf}{} 

\long\def\mklr#1{}
\long\def\mklrc#1{}
\long\def\mklsrc#1{}

\def\arcsec{\hbox{$^{\hbox{\rlap{\hbox{\lower4pt\hbox{$\,\prime\prime$}}}\hbox{$\frown$}}}$}}

\title[MHD-based disc wind prescription]{An MHD-based model for wind-driven disc-planet interactions}

\author[Hammer \& Lin]{Michael Hammer$^{1}$\thanks{E-mail: mhammer@arizona.edu},
Min-Kai Lin$^{1, 2}$ \\
$^{1}$ Institute of Astronomy and Astrophysics, Academia Sinica, Taipei 10617, Taiwan \\
$^{2}$ Physics Division, National Center for Theoretical Sciences, Taipei 10617, Taiwan
}

\begin{document}

\date{Accepted XXX. Received YYY; in original form ZZZ}

\pagerange{\pageref{firstpage}--\pageref{lastpage}} \pubyear{2024}

\maketitle

\label{firstpage}

\begin{abstract}
Hydrodynamic simulations of protoplanetary discs with planets typically assume that the disc is viscously driven, even though magnetic disc winds are now considered the primary driver of angular momentum transport through the disc. Magnetic disc winds are typically left out of hydrodynamic simulations because they require a magneto-hydrodynamic (MHD) treatment and an entire 3D domain, both of which are computationally expensive. Some studies have attempted to incorporate disc winds into disc-planet simulations without full MHD by adding a torque to mimic the effects of a disc wind. However, these studies predate any explicit 3D MHD simulations of planets in the presence of a disc wind. In light of recent MHD studies of disc winds beginning to include a planet, we develop a new disc wind prescription based on these studies and test its efficacy. With three main components, namely (i) excess torque in the planetary gap region, (ii) an MHD-based radial profile for the background torque, and (iii) a moderate level of viscosity, we find that we can essentially reproduce planetary gap profiles for planets above the thermal mass. With lower-mass planets, however, we find it more difficult to reproduce their gap structure. Lastly, we explore the planet's migration path and find that the planet rapidly migrates inwards due to the excess torque in the gap.


\end{abstract}




\section{Introduction} \label{sec:intro}

The hallmark feature of protoplanetary discs in ALMA observations of mm dust is an alternating pattern of gaps and rings spread across the disc \citep[e.g.][]{DSHARP-I}. It is expected that planets are preferentially responsible for carving out many of these gaps, expelling material away from their vicinity as they orbit. Whether a planet is massive enough to open up a gap depends on whether its gravitational torque can compete against the torques associated with the flow of angular momentum transport through the disc. As a result, it is critical for hydrodynamic simulations to accurately represent the flow of angular momentum transport to infer planet and disc properties from observations or to model the planet's orbital migration. 


How do protoplanetary discs transport angular momentum? They do not have an appreciable molecular viscosity. Nevertheless, they must be capable of transporting or losing angular momentum to explain observed stellar accretion rates \citep[e.g.][]{alexander14, hartmann16}. It was long accepted that a turbulent viscosity arising from magneto-hydrodynamic (MHD) effects, namely the magneto-rotational instability \citep[MRI:][]{balbus98}, was responsible. However, MHD studies have recently demonstrated that non-ideal MHD effects suppress the MRI \citep{bai13a, bai17}, particularly in the disc's midplane. The suppression of the MRI gives way for other MHD effects, particularly laminar magnetic stresses and magnetically driven disc winds \citep{bai13a, bai16}, to transport or remove angular momentum, respectively. 

Magnetic winds refer to the magnetically-driven ejection of material from the surface of the disc. The ejected material gets caught along poloidal magnetic field lines and is expelled outwards due to the rotation of the disc \citep{blandford82}. The magnetic field lines responsible thread the midplane and extend to large distances both radially and vertically away from the launching point. The outwards ejection of material and removal of angular momentum gives rise to a corresponding inward flow of material, often primarily near the surface of the disc.

The large radial and vertical extents of the magnetic field lines that govern disc winds require simulations of this phenomenon to have large computational domains in the radial and latitudinal directions, in addition to full MHD as well as the relevant non-ideal MHD effects one would want to include to make the study applicable to protoplanetary discs. Studies that include a planet also require 3D. As a result, 3D MHD studies of planetary gap profiles in wind-driven discs only began very recently \citep{aoyama23, wafflard23, hu25}. Each study, however, only covered a small number of cases over a limited number of orbits. Due to the high computational cost of explicit 3D MHD simulations, it would be advantageous if disc winds could be modeled within a 2D hydrodynamic framework, as is already ubiquitously done with the $\alpha$-viscosity prescription \citep{alpha} for viscous discs. This would make it possible to run larger parameter studies and much longer simulations than are currently feasible with 3D MHD.

Indeed, even before 3D MHD studies were carried out, several recent works attempted to model disc-planet interactions with disc winds without explicit MHD, mainly to study planet migration in 2D \citep{kimmig20} and 3D \citep{lega22}, to study gap profiles \citep{elbakyan22}, or planet-induced vortices \citep{wu23}.\footnote{\cbf{In addition to these 2D studies, there have also been many 1D studies using prescribed wind models that began earlier and have continued to develop \citep{suzuki10, suzuki16, tabone22, okuzumi25}.}} These works incorporated disc winds by prescribing the torque due to the wind, and may also include a mass loss term corresponding to the wind. \cite{kimmig20} and \cite{lega22} found that disc winds altered the co-rotation region enough for planets to uncharacteristically migrate outwards. \cite{elbakyan22} and \cite{wu23} found very different gap profiles compared to wind-less discs\mklrc{compared to windless disks?} in which the outer gap edge ended up much closer to the planet. \cbf{More recent work has continued to use these prescribed models to study planet migration \citep{wu25} and to model the clumpy ring in DM Tau \citep{wu24}.} 

It \cbf{has} remained unclear whether those models and results would hold up against explicit MHD simulations. With such simulations now available, it is of interest \cbf{to check if these} hydrodynamic prescriptions of disc winds \cbf{work, if they can be improved, and to figure out what they can be used for, particularly with} long-term simulations and parameter studies that are still not that feasible with explicit MHD. All three of \cite{aoyama23, wafflard23, hu25} found that winds produce a much stronger torque in \cbf{planetary gaps}, which \cbf{at least early on} carves out a deeper \cbf{gap} at a much faster rate. The outer gap edge was also not noticeably closer to the planet like with prescribed models. Lastly, \cite{aoyama23} found the expected planet migration direction to be inwards, even though the co-rotation region was disrupted in a similar manner to what can happen with a prescribed model. \cbf{On the other hand, \cite{wafflard25} were the first to let the planet migrate and preferentially found slow outward migration.}


\cite{aoyama23} did attempt to fit their 3D MHD gap profiles with 2D hydrodynamics, albeit only with viscous and inviscid models, not any prescribed wind models. They recommended their inviscid model as the better way to mimic 3D MHD. Nonetheless, they did not find either model to be adequate, motivating the need for a prescribed wind model to produce a more optimal fit. 

In this work, we design a new disc wind prescription model and test if it can better match the gap profiles from the MHD simulations with disc winds, focusing on the study by \cite{aoyama23}, hereafter A\&B 2023\mklrc{probably better to alias it to AB23. A: will do if asked in review}, for comparison. We use \cite{kimmig20}, hereafter K20, as a reference wind model to test if and how our model makes a difference. We also use our wind model to explore how the planet migrates in these conditions.

This paper is organized as follows: In Section~\ref{sec:methods}, we describe the setup for our 2D simulations, including the components of our disc wind model and their motivation. We also review the reference prescribed model. In Section~\ref{sec:gap-profiles}, we calibrate our prescribed wind model by matching the resulting gap profile with that found in explicit MHD simulations. We review each component's role and demonstrate that changing the parameter values does not improve the model. In Section~\ref{sec:extensions}, we apply our new prescription model to study disc-planet interaction in wind-driven scenarios not yet explored by explicit MHD simulations, while keeping the planet on a fixed orbit. In Section~\ref{sec:migration}, we apply the model to a migrating planet, exploring cases both with inwards and outwards migration expected. In Section~\ref{sec:discussion}, we discuss the validity of our model and how it could be applied to other studies. In Section~\ref{sec:conclusions}, we conclude our results.

\section{Setup} \label{sec:methods}

We consider a 2D viscous disc of gas and dust harbouring a planet in orbit around a star of mass $M_\bigstar$ at the center of the system. The gas is subject to the effects of a disc wind, namely an inward torque driven by the wind and mass lost to the wind.

\subsection{Disc model} \label{sec:disc}

The disc is set in a cylindrical polar coordinate system ($r, \phi, z$) centered at the star. With the disc's thin structure, we assume the razor thin disc approximation and treat the disc as 2D, where the 2D state variables are vertical averages of their 3D counterparts. \mklrc{include z dimension here because its needed in describing MHD winds in 2.3.1. then go to razor thin disk for the following} The disc's evolution in time $t$ is dictated by the Navier-Stokes equations, specifically the continuity equation and the momentum equation. The gas continuity equation governing the evolution of the surface density $\Sigma$ is 
\begin{equation} \label{eqn:continuity}
\frac{\partial \Sigma}{\partial t} + \vec \nabla \cdot (\Sigma \vec{v}) = -\dot{\Sigma}_\mathrm{wind},
\end{equation}
where $\vec v$ is the velocity vector and $\dot{\Sigma}_\mathrm{wind}$ is the mass loss term due to the wind. The momentum equation governing the velocity is
\begin{align} \label{eqn:navier-stokes}
\Sigma \Big( \frac{\partial \vec{v}}{\partial t} + \vec{v} \cdot \vec  \nabla \vec{v} \Big) = &- \vec \nabla P - \Sigma \vec \nabla \Phi + \vec  \nabla \cdot \overset{\text{\tiny$\leftrightarrow$}} T - f_\mathrm{wind}\hat{\phi} \\
&-  \Sigma [2 \vec \Omega_\mathrm{f} \times \vec v + \vec \Omega_\mathrm{f} \times \big(\vec \Omega_\mathrm{f} \times \vec r \big) + \dot{ \vec \Omega}_\mathrm{f} \times \vec r ]
\nonumber
\end{align}
\mklrc{check if f and F are the same. also check the negative sign: Answer: I believe both are correct now. ("f" and "negative")}
where $P$ is the pressure, $\Phi$ is the gravitational potential, $\vec \Omega_\mathrm{f}$ is the angular frequency vector of the reference frame, $f_\mathrm{wind}$ is the force due to the torque from the wind, and $\mathcal{T}$ is the stress tensor. The gravitational potential $\Phi$ is the sum of the star's potential $\Phi_\mathrm{\bigstar}$, the planet's potential $\Phi_\mathrm{p}$, and the indirect potential $\Phi_\mathrm{i}$ arising from the non-inertial reference frame. The stress tensor is given by
\begin{equation} \label{eqn:stress-tensor}
\overset{\text{\tiny$\leftrightarrow$}}{\cal{T}} = \Sigma \nu \Big[ \vec{v} + (\vec{v})^T - \frac{2}{3}(\vec \nabla \cdot \vec{v}) \vec{I} \Big],
\end{equation}
\mklrc{removed the extra Sigma (typo) in the transpose of the velocity}where $\nu$ is the kinematic viscosity of the disk and $\vec{I}$ is the identity tensor. The disc is assumed to be locally isothermal\mklrc{list EOS, P = rho cs squared and specify the cs power law}, neglecting the energy equation, to match the MHD simulations with disc winds that we wish to compare to. As such, the pressure relates to the density as $P = \Sigma c_\mathrm{s}^2$, where $c_\mathrm{s} = H \Omega$ is the sound speed, $H$ is the disc scale height, and $\Omega$ is the orbital frequency. The disc aspect ratio is kept flat -- in other words, with no flaring -- such that $h~=~H/r$. The orbital frequency is the Keplerian orbital frequency $\Omega \approx \Omega_\mathrm{K} = \sqrt{GM_\bigstar/r^3}$, where $G$ is the gravitational constant. As such, the sound speed scales as $c_\mathrm{s} \propto r^{-1/2}$. Our fiducial setup uses a constant $\nu$ throughout the disc, but some alternate setups use the standard $\alpha$-viscosity model \citep{alpha}, where $\nu = \alpha c_\mathrm{s} H$.

The dust is small enough to be treated as a fluid. It behaves according to the same equations as the gas, except that it is affected by drag forces between the gas and the dust instead of pressure or viscosity. The additional drag force terms in the momentum equation are:
\begin{equation} \label{eqn:rad_drift}
\left.\frac{\partial v_\mathrm{r, d}}{\partial t}\right|_\mathrm{drag} = - \frac{v_\mathrm{r, d} - v_\mathrm{r}}{t_\mathrm{s}},
\end{equation}
\begin{equation} \label{eqn:az_drift}
\left.\frac{\partial v_\mathrm{\phi, d}}{\partial t}\right|_\mathrm{drag} = - \frac{v_\mathrm{\phi, d} - v_\mathrm{\phi}}{t_\mathrm{s}},
\end{equation}
where $v_\mathrm{r}$ and $v_\mathrm{\phi}$\mklrc{fix notation: phi or theta for the azimuth} are the radial and azimuthal velocity components of the gas, and $v_\mathrm{r,d}$ and $v_\mathrm{\phi,d}$ are the radial and azimuthal velocity components of the dust. The stopping time is defined in the midplane as 
\begin{equation} \label{eqn:stopping}
t_\mathrm{s} = \frac{\mathrm{St}}{\Omega} = \left( \frac{\pi}{2} \frac{\rho_\mathrm{d} s}{\Sigma} \right) \frac{1}{\Omega},
\end{equation}
where $\rho_\mathrm{d} = 1$ g / cm$^3$ is the physical density of each dust grain, and $s$ is the size of each grain. Because of the small size of the grains near millimeter-size, the stopping time is in the Epstein regime \citep{weidenschilling77}. Its dimensionless form is the Stokes number St, and we consider grains with fixed St. \mklrc{what are the grain sizes and Stokes numbers we consider? do we fix s or St?}

Beyond drag forces, another effect only experienced by the dust is diffusion, which is part of the dust continuity equation and defined as
\begin{equation} \label{eqn:diffusion}
\left.\frac{\partial \Sigma_\mathrm{d}}{\partial t}\right|_\mathrm{diff} = \nabla \cdot \left( D \Sigma_\mathrm{tot} \nabla\left( \frac{\Sigma_\mathrm{d}}{\Sigma_\mathrm{tot}}\right) \right),
\end{equation}
where $\Sigma_\mathrm{d}$ is the dust surface density, $\Sigma_\mathrm{tot} = \Sigma + \Sigma_\mathrm{d}$ is the total surface density, and the diffusion coefficient is $D = \hat{D} r_\mathrm{p}^2 \Omega_\mathrm{p}$ with dimensions in terms of the the planet's semi-major axis $r_\mathrm{p}$ and the planet's orbital timescale $\Omega_\mathrm{p}$. 
\mklrc{define subscript p} The dimensionless diffusion is on the order of the dimensionless viscosity, i.e. $\hat{D} \approx \hat{\nu}$, since the dust grains are small enough to satisfy St$~\ll~1$ \citep{youdin07}.

\subsection{Planet model} \label{sec:planet}

We incorporate the planet as a gravitational potential according to
\begin{equation} \label{eqn:potential}
\Phi_\mathrm{p}(\mathbf{r}, t) = - \frac{GM_\mathrm{p}}{\sqrt{(\mathbf{r} - {\mathbf{r}_\mathrm{\mathbf{p}}})^2 + r_\mathrm{s}^2}},
\end{equation}
where $M_\mathrm{p}$ is the planet's mass, $r_\mathrm{s} = 0.6H$ is the planet's smoothing length\mklrc{what do we use?}. We also include the indirect potential $\Phi_\mathrm{i}(\mathbf{r}, t) = -GM_\mathrm{p}~[\mathbf{r}~\cdot~\mathbf{r_\mathrm{p}}] / r_\mathrm{p}^3$ arising from the non-inertial reference frame centered on the star. For convenience, we denote the planet-to-star mass ratio as $q$. The planet's gravitational sphere of influence, and the width of the gap in particular, scale with its Hill radius $r_\mathrm{H}~=~r_\mathrm{p}~(q/3)^{1/3}$.


\subsection{Disc wind models} \label{sec:discwind}

We model the magnetically-driven disc wind without any explicit magnetic fields by incorporating its two primary effects, a torque that drives inwards accretion and mass loss due to the wind. First, we summarize the behavior of these effects in MHD\mklrc{no need to keep saying actual or explicit} simulations along with the relevant theoretical background in Section~\ref{sec:background} and then outline the specifics of our implementation in  Section~\ref{sec:prescription}. For reference, we compare our model to that developed by K20, which we review in Section~\ref{sec:reference}.

\subsubsection{Background} \label{sec:background}

\mklrc{need a figure here showing the different torque (MHD, kimmig/yinhao's, this work. would a table comparing the different prescriptions for Sigmawind and Fwind be helpful? A: Added the figure, but didn't show Kimmig's for comparison (should add later!). Didn't add the table.}

With MHD in effect, the wind drives accretion through the conservation of angular momentum. The wind ejects gas from a few scale heights above the midplane up and out of the disc, and this outflow in turn exerts a torque on the disc beneath it. Through the balance of the Lorentz force and the Coriolis force, that torque drives accretion.

\textit{Where does the accretion take place?} The amplitude of the wind-driven accretion at different heights is related to the balance of those forces through
\begin{equation} \label{eqn:real-accretion-velocity}
\frac{1}{2} \rho \Omega_\mathrm{K} v_r \approx \frac{B_\mathrm{z}}{4 \pi} \frac{\partial B_\phi}{\partial z},
\end{equation}
where $\rho$ is the 3D gas density, $\Omega_\mathrm{K}$ is the Keplerian velocity\mklrc{do we need to distinguish between OmegaK and Omega? I assume there is only one Omega (which is OmegaK) in the preceding sections pertaining to sound-speed profiles and for setting the stopping time. A: They are all roughly the same, but isn't it more appropriate to use the value here instead of the variable?}, and $B_\mathrm{\phi}$ and $B_\mathrm{z}$ are the azimuthal and vertical components of the magnetic field \citep{wardle07, bai13a}. \mklrc{eq 9 is in cgs units? A: I think so. Is that an issue?} As a result, the wind-driven accretion is stronger at heights where there is a strong vertical gradient in $B_\phi$, which typically occurs where is a sign flip in the magnetic field. The general vertical accretion profile can vary. It could be spread across a range of heights from the midplane to the wind base \citep[e.g. Figure 9 in][]{cui21}. It may instead be confined to a relatively thin current layer near the base of the wind if the field lines near the midplane are too straight and the field strength is constant due to non-ideal diffusive effects \citep[e.g. Figures 10 and 11 in][]{bai13a}. \mklrc{better to cite recent global models as these have superseded previous generation of shearing box sims. A: Do you have suggestions?} The accretion can happen on both sides of the midplane, or only above it in either case.

\textit{How does the wind-driven accretion compare to radially-driven accretion such as viscosity?} The corresponding vertically-integrated accretion rate induced by the wind torque is
\begin{equation} \label{eqn:real-wind-accretion}
\dot{M} = \frac{8 \pi}{\Omega} r |\mathcal{T}^\mathrm{Max}_\mathrm{z \phi}|_\mathrm{z_b},
\end{equation}
\cbf{where $\mathcal{T}^\mathrm{Max}_\mathrm{z \phi} = -B_\mathrm{z}  B_\mathrm{\phi}$} is the vertical component of the Maxwell stress tensor evaluated at the base of the wind $z = z_\mathrm{b}$ \cbf{and the wind is assumed to be symmetric about the midplane}. The level of accretion associated with this Maxwell stress is a factor of $h^{-1}$ stronger than the rate associated with a radially-driven stress $\mathcal{T}_\mathrm{r \phi}$ of the same magnitude. \cbf{That radial stress could be either the Reynolds stress $\mathcal{T}^\mathrm{Rey}_\mathrm{r \phi} = \delta (\Sigma v_\mathrm{r} )  \delta v_\mathrm{\phi}$ or the Maxwell stress $\mathcal{T}^\mathrm{Max}_\mathrm{r \phi} = -B_\mathrm{r}  B_\mathrm{\phi}$.} The quantity $r \mathcal{T}^\mathrm{Max}_\mathrm{z \phi}$ is the wind torque itself. \mklrc{what units are we using?}

\textit{Which torques are associated with the wind?} The torque derives from the angular momentum flux associated with the wind. A\&B 2023 emphasize that this angular momentum flux involves two components, the cumulative wind torque itself $\Gamma_\mathrm{wind}$ and the angular momentum flux $J_\mathrm{Max}$ associated with the horizontal Maxwell stress $T^\mathrm{Max}_\mathrm{r \phi}$. These vertical and horizontal components are not truly separate, and are both just components of the poloidal Maxwell stress split for the sake of measuring the fluxes. As such, they can be combined into a single MHD angular momentum flux where $J_\mathrm{MHD} = J_\mathrm{Max} + \Gamma_\mathrm{wind}$. The corresponding MHD torque is then just $dJ_\mathrm{MHD}/dr$. \mklrc{do the units match? J is a flux, T is a torque}

\textit{What is the radial profile of the torque associated with the wind?} The profile for the MHD torque $dJ_\mathrm{MHD}/dr$ in A\&B 2023 \mklrc{A\&B2023} is piecewise split into two main regions, one part corresponding to the gap and the rest corresponding to the background away from the gap, as illustrated in their Figure 14. The background torque in the outer disc in particular follows a $r^{-1/2}$ profile. Inside the gap region, the torque is amplified by a factor of about \cbf{three to} five. This increase is due to the lower density in the gap increasing the magnetic flux concentration, \cbf{which in turn also results in a higher torque (see Figure 3 in \citealp{bai16} or \citealp{lesur21}}). The profile in this region is roughly constant, with a magnitude of about $2.3 \times 10^{-3}~r_\mathrm{p}^2 \Sigma_\mathrm{p} v_\mathrm{p} \Omega_\mathrm{p}$ regardless of planet mass, where $\Sigma_\mathrm{p}$ is the initial surface density at the location of the planet and $v_\mathrm{p}$ is the velocity of the planet. \mklrc{define vp} The width of the gap region scales roughly with the Hill radius, and is about $1.5$ to $2.0~r_\mathrm{H}$. \mklrc{has RHill been defined? can define it in planet potential section. A: done.} It's less clear what the behavior of the torque in the inner disc should be since a pure MHD-driven gap \citep[e.g.][]{riols20} forms in all cases at what is considered to be a random location.

\textit{What is the radial profile of the accretion rate associated with the wind?} The background wind-driven accretion rate has a loosely constant profile in A\&B 2023, as illustrated in their Figure 9. The precise power law is slightly negative. Because of the gap's excess torque, each case also has a bump in accretion in the gap, which has a decreasing amplitude with increasing planet mass. In the outer part of the disc away from the planet at $r > 1.5~r_\mathrm{p}$, the accretion profile begins to change. In their two cases with lower-mass planets, the accretion starts increasing with radius. Meanwhile in the highest-mass case, it drops off completely before increasing again like the other cases. \mklrc{question: a non-constant Mdot implies unsteady state. if our prescription induces a non-uniform Mdot, how do our runs reach steady state? is it just compensated by mass loss? A: they do not reach steady state.}

\textit{What is the radial profile of the mass lost through the wind?} The profile for the wind mass loss in A\&B 2023 is highly non-uniform, as illustrated in their Figure 9. The expected ratio of the wind mass loss rate to the wind-driven accretion rate, \cbf{assuming a vertically-dominated outflow}, is given as
\begin{equation} \label{eqn:masslossratio}
\frac{1}{\dot{M}_\mathrm{acc}} \frac{d \dot{M}_\mathrm{wind}}{d \ln r} \approx \frac{1}{2} \frac{1}{(r_\mathrm{A} / r_\mathrm{wb})^2 - 1},
\end{equation}
where \cbf{$\lambda \equiv (r_\mathrm{A} / r_\mathrm{wb})^2 > 1$} is the magnetic lever arm, $r_\mathrm{A}$ is the Alfv\'en radius where the local velocity is the Alfv\'en velocity $v_\mathrm{A} = B / \sqrt{4 \pi \rho}$, and $r_\mathrm{wb}$ is the location of the base of the wind \citep{ferreira95, bai16}. However, this quantity is not useful near the planet. As a result of there being no poloidal field lines in the gap, mass loss only takes place outside of the planetary gap. In both the interior and exterior regions of the disc, the mass loss rate peaks close to the peak in density at the gap edge, indicating a loose dependence on density between the gap edges. In the gap itself, there is instead a small inflow at a much lower magnitude than the adjacent mass loss.

\textit{Is there any turbulence in the disc?} Although the disc wind primarily induces laminar accretion, the disc can still have a lower level of turbulence through other magnetic effects such as MRI. Turbulence can manifest through the horizontal Reynolds stress $\mathcal{T}^\mathrm{Rey}_\mathrm{r \phi}$ and the horizontal Maxwell stress $\mathcal{T}^\mathrm{Max}_\mathrm{r \phi}$, which each correspond to a dimensionless $\alpha \equiv \mathcal{T}_\mathrm{r \phi} / P$  \citep{alpha}. Figure 7 of A\&B 2023 shows that $\alpha^\mathrm{Max}$ is easily dominant over $\alpha^\mathrm{Rey}$ by about an order of magnitude in both the gap and outer disc. \mklrc{need to define alphaRey and alphaMax. but if they're only used here then you can just say that Maxwell stresses dominate over Reynolds stresses. A: I believe they are defined in the previous sentence.} Both components are much stronger in the gap, and sharply decay in the outer disc but never flatten out.

\subsubsection{New prescribed model} \label{sec:prescription}

We sought to develop a disc wind prescription motivated by the underlying physics of disc winds while focusing on reproducing the gap and outer disc profiles obtained from explicit MHD simulations of planets in wind-driven discs. We do not attempt to model the inner disc due to the purely MHD-driven gaps that form there (A\&B 2023). Our goal is to derive a simple prescription to study different parameters and longer timescales that MHD simulations have not yet explored. \mklrc{this is the fiducial setup, but further, minor modifications are made in sections blah}

This model is tailored towards fitting the MHD gap profiles on the short timescales we have. For the longer-term simulations in the application sections with both a static planet (Section~\ref{sec:longterm}) and a migrating planet (Section~\ref{sec:migration}), we \cbf{also} use the simpler mass loss prescription \cbf{from K20}, since our fiducial mass loss profile is tailored towards short-term simulations only.

\paragraph{Torque prescription} \label{sec:torque}

\begin{figure} 
\centering
\includegraphics[width=0.47\textwidth]{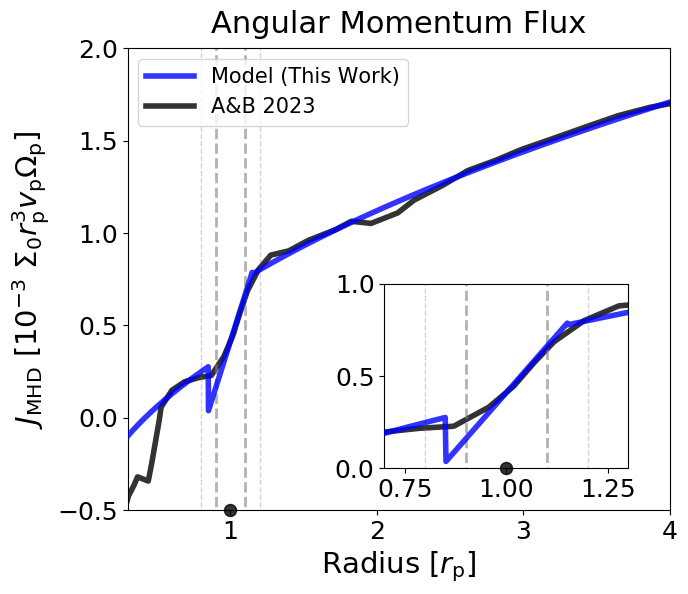} \\
\vspace{2em}
\includegraphics[width=0.47\textwidth]{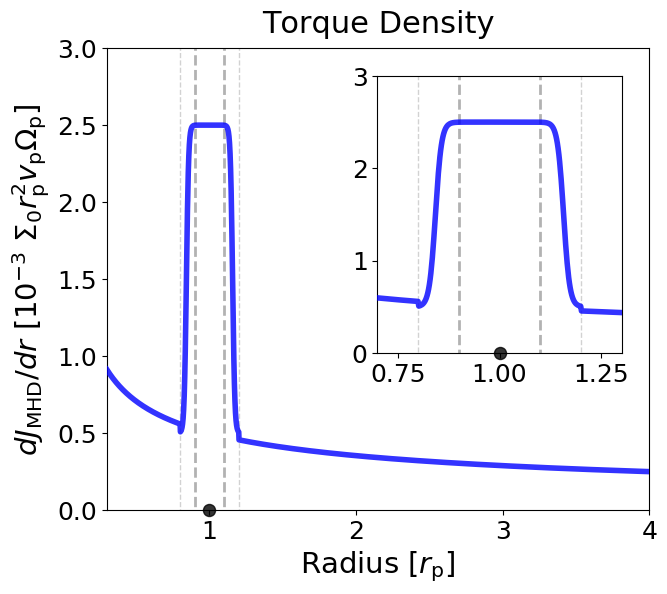}
\caption{The prescribed torque is based on the angular momentum flux $J_\mathrm{MHD}$ (\textit{top}) and the resulting torque density  $dJ_\mathrm{MHD}/dr$ (\textit{bottom}). Our prescribed model (\textit{blue}) is intended to match the actual MHD simulation (\textit{black}) by \citealp{aoyama23} in the gap and the outer region of the disc. References lines at $\pm 1.0~r_\mathrm{H}$ and $\pm 2.0~r_\mathrm{H}$ are included (\textit{grey dashed lines}) to help demarcate the different regions, while the planet is indicated with a dot. We include the smoothing of $dJ_\mathrm{MHD}/dr$ in the plot, which affects the profile mainly at $1.4~r_\mathrm{H} \le | r - r_\mathrm{p}| \le 1.85~r_\mathrm{H}$, but leave out the corresponding smoothing of $J$.} 
\label{fig:torque-profile}
\end{figure}

Following the largely piecewise torque split in A\&B 2023, we prescribe the torque with two largely piecewise components, one for the background and one for the gap, and a smoothing between the two components, \cbf{as depicted in Figure~\ref{fig:torque-profile}}. We express both components of the vertically-integrated azimuthal forcing in the momentum equation with the form
\begin{equation} \label{eqn:wind}
f_\mathrm{wind} = \frac{1}{2 \pi r^2 \Sigma} \frac{dJ_\mathrm{MHD}}{dr}
\end{equation}
We adopt this form based on Equation 14 from A\&B 2023.

We set the MHD torque to be a power law based on the torque profiles from A\&B 2023 to be
\begin{equation} \label{eqn:J_MHD}
\frac{d J_\mathrm{MHD}}{d r} = K b \hat{r}^{n} ~  r_\mathrm{p}^2 \Sigma_\mathrm{p} v_\mathrm{p} \Omega_\mathrm{p},
\end{equation}
where $\hat{r} = r / r_\mathrm{p}$, $b$ is the torque coefficient, and both the gap enhancement factor $K(r)$ and the power $n(r)$ vary with radius depending on whether a location is in the gap or not.\mklrc{should it be Sigmap instead of Sigma0?} Specifically, $K$ has the form 
\begin{equation} \label{eqn:smoothing}
K(r) = 1 + \frac{1}{2} (K_\mathrm{gap} - 1) \Bigg[1 - \tanh \Big( \frac{\Delta r - r_\mathrm{t}}{ r_\mathrm{w}}  \Big) \Bigg],
\end{equation}
where the smoothing is determined by two parameters: the transition radius $r_\mathrm{t}$ and the transition width $r_\mathrm{w}$, while the distance from the planet is  $\Delta r \equiv |r - r_\mathrm{p}|$. \mklrc{need to define sp. is it necessary to use the normalized distance s? seems inconvenient to normalize the global cylindrical radius by the local Hill radius. in the tanh function, you could equally use r-rp, rt, and some width w, since the scalings cancel out} 
Inside the gap, $K \to K_\mathrm{gap}>1$, while outside the gap there is no enhancement and $K \to 1$. We set the transition width to be $r_\mathrm{w} = 0.15~r_\mathrm{H}$ and the transition radius to be $r_\mathrm{t} = 1.625~r_\mathrm{H}$, which is the gap cutoff of $1.5~r_\mathrm{H}$ plus half the transition width. (A location outside the gap cutoff is chosen so that the decline in the enhancement factor begins at the edge of the gap, not in it.) On the other hand, the power $n(r)$ is simply defined piecewise as a constant-slope $n = 0$ in the gap and $n = -1/2$ outside the gap. Overall, the full form in the two main regions tends to
\begin{equation} \label{eqn:J_MHD}
\frac{d J_\mathrm{MHD}}{dr} \to r_\mathrm{p}^2 \Sigma_\mathrm{p} v_\mathrm{p} \Omega_\mathrm{p}
  \begin{cases}
    K_\mathrm{gap} b & \text{inside the gap} \\
    b \hat{r}^{-1/2} & \text{outside the gap}, 
  \end{cases}
\end{equation}
where ``inside-the-gap" is $\Delta r < 1.5~r_\mathrm{H}$, ``outside-the-gap" is $\Delta r > 2.0~r_\mathrm{H}$, and the form in the smoothing region ($1.5~r_\mathrm{H} < \Delta r < 2.0~r_\mathrm{H}$) is more complex due to the smoothing of $K$. We neglected to smooth $n$ because it did not contribute as much to the difference between the magnitude of the torque inside and outside the gap, and would make the model unnecessarily more complicated. 

\mklrc{what is actually applied in the code? eq 13 with K given by eq 14 and n depending on inside/outside gap; or eq 13 with 15? if eq 15 is only intended to show the  limiting for of dJ/dr inside/outside the gap; then use the "tend to" notation. but if eq 15 is the one actually implemented in the code, then no need to show eq 14. A: changed to "tend to"}

The two physical parameters are set as $b = 5.0 \times 10^{-4}$ and $K_\mathrm{gap} = 5$ so that $Kb = 2.5 \times 10^{-3}$ in the gap. This gap value is intended to approximately match the value of $dJ_\mathrm{MHD}/ dr \approx 2.3 \times 10^{-3}~r_\mathrm{p}^2 \Sigma_\mathrm{p} v_\mathrm{p} \Omega_\mathrm{p}$ that A\&B 2023 find in the gap in all cases regardless of planet mass. \cbf{The value of $K_\mathrm{gap} = 5$ is close to the maximum value observed by A\&B 2023 over time. We explain in Section~\ref{sec:gap-evolution} why we never use a lower value, even early on when the torque amplification in the gap is lower in the MHD simulations.}

The prescribed torque drives mass accretion through the disc. 
Neglecting the smoothing factor, this accretion rate \cbf{initially} is
\begin{equation} \label{eqn:mass-accretion}
\dot{M}(r) = \frac{2}{r \Omega} \frac{dJ_\mathrm{MHD}}{dr} = 2b ~r_\mathrm{p}^2 \Sigma_\mathrm{p} \Omega_\mathrm{p}
  \begin{cases}
    K \hat{r}^{1/2} & \text{inside gap} \\
    1 & \text{outside gap}
  \end{cases}
\end{equation}
in the two main regions of the disc. With just this radial dependence of the torque in the model, the accretion would be in a steady state outside of the gap. From inside the gap, there would be excess flow of gas into the inner disc regardless of gap depletion. \mklrc{comment on non-steady Mdot inside gap. this implies a mass loss from the gap region. A: Done, but double check the lack of dependence on surface density.}

The above accretion rate corresponds to an inwards radial velocity of 
\begin{equation} \label{eqn:vr}
\frac{v_\mathrm{r}(r)}{v_\mathrm{p}} = -\frac{1}{v_\mathrm{p}} \frac{2 f_\mathrm{wind}}{\Omega} = -\frac{b}{\pi} \frac{\Sigma_\mathrm{p}}{\Sigma} \frac{ \Omega_\mathrm{p}}{\Omega}
  \begin{cases}
    K \hat{r}^{-2} & \text{inside the gap} \\
    \hat{r}^{-5/2}& \text{outside the gap}.
  \end{cases}
\end{equation}
\mklrc{consider flipping order of vr and Mdot}
With the surface density power law of $\Sigma \propto r^{-1.25}$ from A\&B 2023, this radial velocity has a negative power law everywhere in the disc, but is closer to constant outside the gap at $v_\mathrm{r} \propto r^{-0.25}$. \mklrc{fix/restore units for clarity}


\paragraph{Mass loss prescription} \label{sec:massloss}

Based on the MHD mass loss profile, we express the mass loss term in the continuity equation as
\begin{equation} \label{eqn:massloss}
\dot{\Sigma}_\mathrm{wind} = \bar{b} \frac{\Omega_\mathrm{p}}{2 \pi}\Sigma,
\end{equation}
where $\bar{b}$ is the mass loss rate. The mass loss rate is equivalent to the torque coefficient $b$ away from the planet, but \cbf{is} reduced to zero near the planet according to
\begin{equation} \label{eqn:b}
\bar{b} = 
  \begin{cases}
    b & \text{if } |r - r_\mathrm{p}| > 1.0~r_\mathrm{H} \\ 
    0 & \text{otherwise}
  \end{cases}
\end{equation}
where $r_\mathrm{H}$ is planet's Hill radius. The mass loss is reduced to zero in the planet gap to reflect that there is no mass loss in that region. The surface density dependence is intended to capture the mass loss rate peaking near the location with the highest density at each gap edge\mklrc{rephrase "peaking at the peaks"}, and also helps smooth the discontinuity between the two regions. We neglect the small mass inflow in the gap because we find that the gap already depletes too slowly compared to the  MHD simulations. We also neglect that the mass loss rate outside of the gap is non-uniform and not exactly proportional to the surface density because our main focus \cbf{is} to approximate the profile the best near the outer gap edge.

\paragraph{Viscosity prescription} \label{sec:viscosity}

Besides including the two main effects of the wind, we also account for turbulence in the disc. Even though the level of turbulence should not be too important for accretion relative to the wind, we anticipate it may have a stronger effect on the gap profile. 

We presume that the Maxwell stress $\mathcal{T}^\mathrm{Max}_\mathrm{r \phi}$ has two components, a primary component that induces laminar accretion and a secondary component that induces a low amount of turbulence. To keep the turbulent component of the overall accretion small, we prescribe a small constant $\nu = 10^{-5}~r_\mathrm{p}^2 \Omega_\mathrm{p}$. For simplicity, we drop the units of $\nu$ hereafter.\mklrc{just use nu hat. A: normally, I drop the hat. I think that's more consistent with its counterpart "b".} The viscosity is kept constant to keep the model simple and to make it easier to compare to viscous studies \cbf{that} use a constant viscosity. We note that although $\mathcal{T}_\mathrm{z \phi}$ drives more accretion than an $\mathcal{T}_\mathrm{r \phi}$ of equal magnitude, that extra factor has already been absorbed into $b$. In our model, the wind-driven and viscous accretion rates relate through the ratio $\dot{M}_\mathrm{b} / \dot{M}_\mathrm{\nu} = (b / \nu) / \pi$.





\subsubsection{Reference prescribed model} \label{sec:reference}

To understand if our model and its components have any effect on the gap profiles, we compare our model to the one developed by K20, which contains two components: a torque prescription and a mass loss prescription.


The torque term in the K20 model is given as
\begin{equation} \label{eqn:kimmig-torque}
\Gamma = \dot{\Sigma}_\mathrm{wind} \Omega_\mathrm{K} r^2 (\lambda - 1),
\end{equation}
taken from their Equation 5. The lever arm $\lambda$ is set to their fiducial value of $\lambda = 2.25$. There are two key differences. The main one is the resulting power law in the radial direction of the torque. Their torque follows a $\Gamma \propto r^{-2.25}$ profile \cbf{with our fiducial surface density profile (see Section~\ref{sec:simulations})}, whereas the torque in our model follows a flatter $\Gamma \propto r^{-1.5}$ profile. 
A more minor difference is that we do not include the lever arm.


The mass loss term in the K20 model is 
\begin{equation} \label{eqn:ref-massloss}
\dot{\Sigma}_\mathrm{wind} = b \frac{\Omega_\mathrm{K}}{2 \pi}\Sigma,
\end{equation}
taken from their Equation 3, where $b$ serves the same purpose and is the basis for our definition of $b$. \mklrc{is b and b bar the same?. A: Yes.} There are two slight differences. First, they apply the mass loss everywhere, whereas we only apply it outside the gap. Second, their mass loss rate depends on the local orbital period\mklrc{make it obvious, OmegaK(r). There are many Omegas in the paper: Omega, Omegap, Omega hat, OmegaK. consider simplifying, if possible}, which we neglect in order to boost the mass loss rate at the \cbf{outer} pressure bump. 

The orbital dependence in the mass loss term does not make a significant difference in short-term simulations around several hundred orbits or less, but it does in long-term simulations around 1000 orbits or more. In particular, the outer disc depletes much more slowly away from the gap with the K20 prescription. Since their prescription is better physically-motivated in the longterm, we do include the orbital dependence in our long-term simulations, as we discuss in Section~\ref{sec:longterm}.

Lastly, in some cases, we also add \cbf{into their model} the viscosity and extra gap torque components from our model for comparison purposes.

\subsection{Simulations} \label{sec:simulations}

All of our simulations are run with the FARGO3D hydrodynamic code \citep{FARGO3D, FARGO3D-DUST}. In our fiducial runs, we carry out 2D simulations generally following the setup choices from A\&B 2023. For that reason, the gas surface density distribution follows a power law of $\Sigma_0(r) = \Sigma_p (r / r_\mathrm{p})^{-p}$, the initial surface density at the location of the planet is $\Sigma_\mathrm{p} = 2.315 \times 10^{-4}~M_{\bigstar} r_\mathrm{p}^{-2}$\mklrc{give units or state choice of computational units first}, the surface density power law is $p = 1.25$, and the disc has a flat aspect ratio of $h = 0.1$. The initial dust profile matches the gas power law, except with a 1-to-100 dust-to-gas ratio. Nevertheless, as we neglect self-gravity, our fiducial simulations are scale-free with respect to the density for both the gas and dust.\mklrc{only scale free because there's no self-gravity} The only set of cases in which the density level can affect the results is with a migrating planet. 

The dust Stokes number is set to $\mathrm{St} = 0.023$, which is chosen as an intermediate grain size to monitor the appearance of the dust in vortices. \mklrc{do not mix up grain size and stokes number. rephrase. (do we fix grain size or St?)} \cbf{If the planet is placed at $r_\mathrm{p} = 20~$AU, this Stokes number corresponds to a size of $s = 0.75~\mathrm{mm}~(\Sigma/\Sigma_\mathrm{p})(r / r_\mathrm{p})^{-2}$.} For our main purpose of studying the gap profiles, the dust is not used.

Our fiducial simulations have an arithmetic grid spanning from $r \in [0.3, 4.0]r_\mathrm{p}$ = [$r_\mathrm{in}$, $r_\mathrm{out}$] in radius and full-circle in azimuth $\phi$. The domain has 768 grid cells in both directions. With this radial resolution, the scale height of $H = 0.1~r_\mathrm{p}$ at the location of the planet is resolved by 21 grid cells, about the same as the 20-grid-cell maximum resolution used by A\&B 2023. Such a resolution should be sufficient for studying the gap profiles. However, a more in-depth study focusing on the vortices (and the dust in particular) may require a higher resolution. We chose a relatively low resolution both to match the MHD study and to test the efficacy of our model with an efficient setup that does not\mklrc{i'd avoid abbreviations to make it formal (e.g. doesn't -> does not)} require extensive computational time.

At the boundary, we have wave-killing zones, known in FARGO3D as Stockholm boundary conditions \citep[e.g.][]{deValBorro06}. The inner boundary zone exists from $0.3~r_\mathrm{p} < r  < 0.375~ r_\mathrm{p}$, with the end at $1.25~r_\mathrm{in}$. The outer boundary zone exists from $ 3.36~r_\mathrm{p} < r < 4.0 ~r_\mathrm{p}$, with the start at $0.84~ r_\mathrm{out}$. The damping timescales in both regions are $\tau = 0.3~\Omega^{-1}$. We found that our fiducial results with the prescribed wind are not heavily dependent on the boundary conditions. \mklrc{state the damping rate or time} On the other hand, the gap profiles can change by a significant amount without wave-killing zones when there is no prescribed wind.


\section{Matching MHD Gap Profiles} \label{sec:gap-profiles}

\mklrc{would a table summary of the runs be helpful? A: not going to do.}

With a four-component prescribed disc wind model, we can reasonably match the 3D MHD gap profiles using a 2D hydrodynamic model for cases with a high-mass planet above the thermal mass and the fiducial high aspect ratio of $h = 0.1$. \mklrc{how massive and how high h?} Our prescribed model consists of
\begin{enumerate}
  \item a background torque power-law profile,
  \item a mass loss profile,
  \item viscosity, and
  \item enhanced torque in the planet gap.
\end{enumerate}
As expected, the extra torque in the gap is the key component to deepen the gap enough to match the high gap depths from the  MHD simulations. Meanwhile, the viscosity is the key component to match the gap profile at and around the outer gap edge. In general, the gap-opening process with 2D hydrodynamics is much slower than in 3D MHD. Consequently, we find a better match between the final MHD gap profile from A\&B 2023 at 140 orbits with our 2D hydrodynamic gap at about 210 orbits. 


Our best fit for the gap profiles uses the level of background torque taken from the MHD simulations. On the other hand, the level of viscosity is about one-sixth the level of stress from the MHD simulations, suggesting the Maxwell component that dominates the stress is largely laminar (i.e. five-sixths not turbulent).

There are only a narrow range of values for the viscosity that could be viable. If the viscosity is too high, it would become the main driver of accretion in the disc. It could also prevent the formation of vortices at the outer gap edge, which do appear in the  MHD simulations and fundamentally alter the gap profile. If it is too low, it would let the planet push too much material away from the outer gap edge and prevent the pressure bump from becoming as strong as the ones in MHD.

We were not able to develop as adequate of a fit for the gap profiles induced by lower-mass planets. With lower-mass planets, the pressure bump at the outer gap edge already has too high of a density even with no viscosity. Adding in viscosity only worsens this problem. As such, no amount of constant viscosity results in a good fit.

\mklrc{no figures in this section. is this intentional? A: It's just an intro.}

\subsection{Context} \label{sec:context}

We chose the intermediate-mass case from A\&B 2023 with a 3~Jupiter-mass (3 thermal-mass) planet as our fiducial case. We aim to fit the gap itself and the general outer gap edge profile, focusing on three main quantities: the gap depth, the outer gap edge location (i.e. the gap width or half-width), and the surface density at the outer gap edge.\mklrc{why are these the most important features to reproduce? the reasons may also be highlighted in the main introduction. i.e. we need a more sophisticated wind prescription to get accurate gap structures for blah reason, but without the cost of MHD. A: Added here, but not in the intro.} The last two quantities are measured at the peak in the pressure bump. 

We do not focus on any quantities related to the inner gap edge because A\&B 2023 found a purely MHD-driven gap near there, located at about $r = 0.5~r_\mathrm{p}$ (as shown in their Figure 3, and re-plotted in many of the figures in this work). This additional gap affected the planet's gap structure towards its inner gap edge. Since this feature is not related to the planet and could appear in a different location or not at all in a real disc, we do not attempt to reproduce the inner gap edge and only focus on the gap itself and the outer gap edge.


We expect that if we could match all three quantities, the rest of the profile would also match (aside from the inner gap edge). We chose the location of the pressure bump as the best way to measure the gap width because it can be inferred directly from dust observations. The importance of the gap edge density is that it has some correlation with whether vortices could form through the Rossby Wave instability \citep[e.g.][]{lovelace99, ono16}. We consider the gap depth important because many studies of gap profiles in viscous wind-less discs have focused on developing analytical scaling relations for the gap depth \citep{fung14, duffell15, kanagawa15b}.

A\&B 2023 already attempted to fit their MHD gap profiles with 2D hydrodynamic simulations, but without any prescribed disc wind model. For each of their main three cases (1, 3, \& 5 Jupiter-mass), they ran two comparison cases: one inviscid and one viscous with $\alpha = 6 \times 10^{-3}$,\mklrc{since alpha viscosity is discussed, it needs to be defined somewhere and its related to nu stated. i suggest doing so after giving the viscous stress tensor (eq 3). A: Done, although I needed to be careful about defining alpha after cs} the measured background value in their simulations, which is largely determined by the Maxwell stress. Of the two, they advocate for the inviscid case as the better simple fit, citing its deeper gap and its matching gap width near half-max on the exterior side of the gap. 

Despite those successful facets, neither of the simple fits appeared adequate. Even though the inviscid gap was deeper, both gaps were noticeably too shallow.\mklrc{try to make the writing style more formal. i.e. avoid phrases like "way too...". also make the comments more quantitative, if possible.} Additionally, the gap width in the inviscid case only matched the MHD case along the ``shoulder" of the outer gap edge. However, if one matches the gap width based on the location of the peak in the pressure bump instead --- since this sets the gap width in mm dust observations --- then their viscous case actually performs better than the inviscid case, owing to the inviscid pressure bump spreading out way too far away from the planet.

\mklrc{the last two paragraphs, which summarizes AB23 and highlighting its inadequacies, appears more suitable in introduction as a motivation for this work, i.e. coming up with a new wind prescription. A: Added a short summary to the intro.}


\subsection{Building up the best-fit model} \label{sec:fiducial-case}

In this section, we build up our best-fit model starting out from the two-component prescription with torque and mass loss and then showing how the gap profiles change when adding in viscosity as a third component and extra gap torque as a fourth. At each level of complexity, we compare the gap profiles for both our model and the K20 model to the MHD gap profiles at $t = 140$, the end time of the simulations by A\&B 2023. 
Afterwards, we extend the comparisons of the gap profile to the full evolution over time. We then discuss the roles of the vortices and the inner MHD gap. At the end, we show how varying any of the parameters only worsens the quality of the fit.


\begin{figure} 
\centering
\includegraphics[width=0.47\textwidth]{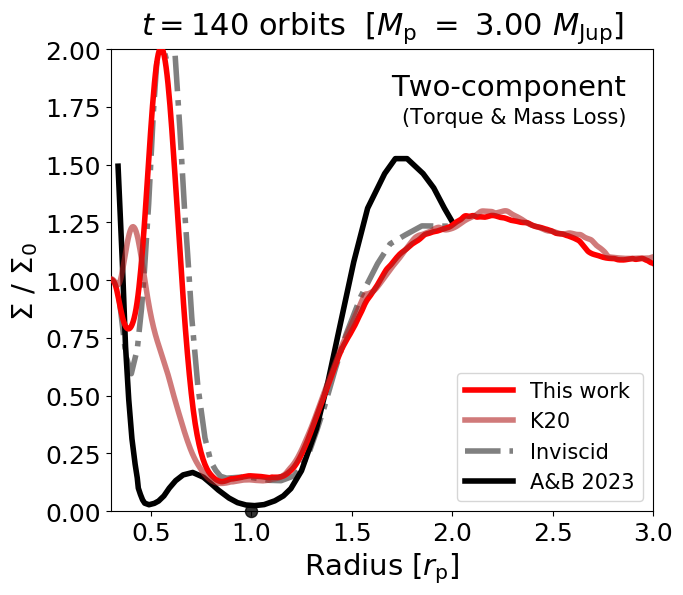}
\caption{Comparison of the two-component prescribed disc wind models from this work (\textit{red}) and by K20 (\textit{light red}) to the  MHD simulation (\textit{black}) by A\&B 2023 and their inviscid comparison run (\textit{grey dashed line}), all at $t = 140$. The two-component models are nearly identical to the inviscid comparison run, and neither does a particularly good job at matching the real MHD gap profile at the trough of the gap or the outer gap edge. The two models themselves are even more identical to each other at $r > 0.8~r_\mathrm{p}$. \textit{Note: The A\&B 2023 profiles are averaged over 10 orbits from $t = 130$ to 140. The profiles from our work are averaged over 2 orbits.}} 
\label{fig:first-comp}
\end{figure}

We begin with the two-component models that consist of a prescribed torque and mass loss.
Figure~\ref{fig:first-comp} compares the gap profiles from the two prescribed models to the MHD gap profile. With just the torque as the main component, the prescribed wind has little effect on the gap profile. The resulting profiles largely match the inviscid comparison run by A\&B 2023, which is not entirely surprising since neither simulation has viscosity. The mass loss component also has little effect on the gap profile exterior to the planet because of the slow dependence on the orbital timescale. As with the inviscid run, the primary outcomes of the two-component model that could be improved are that the amplitude of the pressure bump and the depth of the gap should be increased. 

\begin{figure} 
\centering
\includegraphics[width=0.47\textwidth]{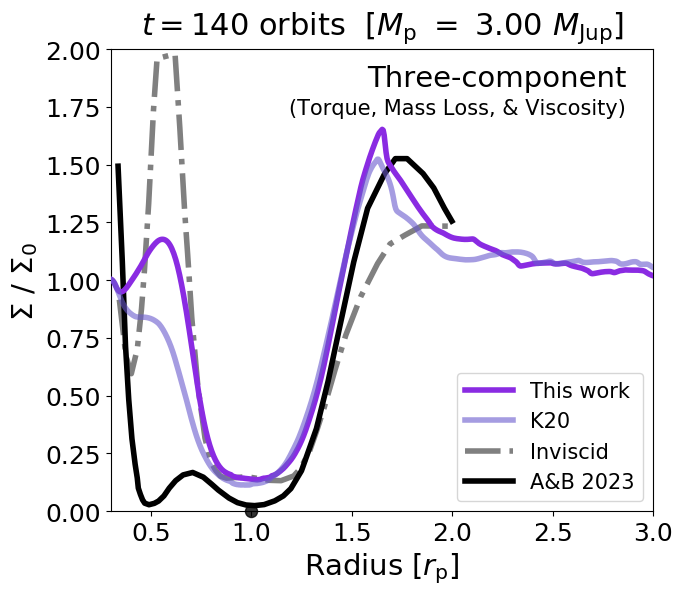}
\caption{Comparison of three-component prescribed disc wind models from this work (\textit{purple}) and by K20 (\textit{light purple}) to the  MHD simulation (\textit{black}) by A\&B 2023 and their inviscid comparison run (\textit{grey dashed line}), all at $t = 140$. The three-component models are much closer to matching the real MHD gap profile than the two-component models, particularly at the outer gap edge.} 
\label{fig:first-viscosity}
\end{figure}

To make the peak of the pressure bump higher, we added viscosity as a third component. 
As Figure~\ref{fig:first-viscosity} shows, incorporating viscosity significantly elevates the pressure bump\mklrc{is this counter-intuitive because viscosity should smooth out bumps? Answer: I already explained this in the next paragraph.}. With a value of $\nu = 10^{-5}$ (equivalent to $\alpha = 10^{-3}$ at $r = r_\mathrm{p}$), not only is the peak of the gap edge much closer to the target location and density from the MHD run, the slope of the inner shoulder of the outer gap edge matches almost perfectly across the entire shoulder. 

The amount of viscosity we use helps match the MHD gap profile in two different ways. First, without any viscosity, there was nothing to oppose the planet driving material away from it. Having viscosity, however, stops that material from spreading out too far from the planet, keeping the pressure bump as strong as it is in the MHD run. Second, our viscous model also works better than the comparison run by A\&B 2023 because we use a lower level of viscosity. That lower amount allows vortices to develop at the gap edge through the Rossby wave instability \citep[RWI:][]{lovelace99, li00, li01}, \mklrc{need a ref for RWI} which also occurs in the MHD run. The viscous comparison run they did was less effective because it used too high of a viscosity, which suppressed the RWI and prevented vortices from ever forming. With the three-component model, the main issue remaining at the outer gap edge is that it is slightly too close to the planet.

\begin{figure} 
\centering
\includegraphics[width=0.47\textwidth]{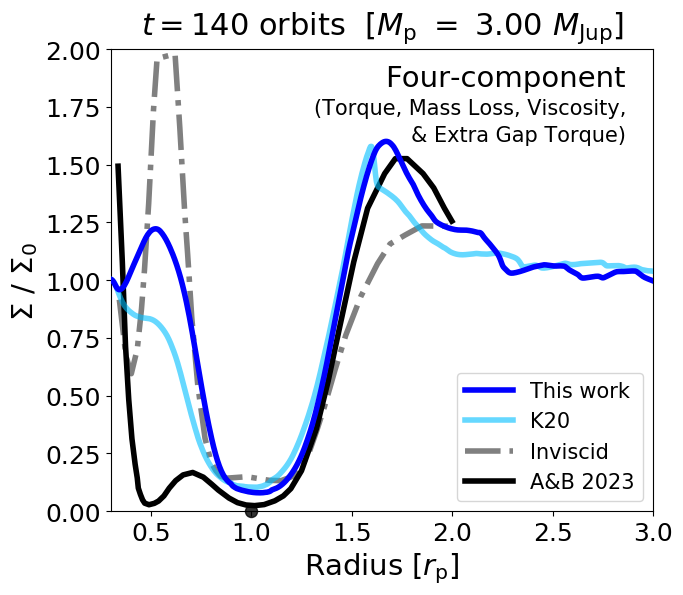}
\caption{Comparison of the four-component prescribed disc wind models from this work (\textit{blue}) and by K20 (\textit{light blue}) to the  MHD simulation (\textit{black}) by A\&B 2023, and their inviscid comparison run (\textit{grey dashed line}), all at $t = 140$. The modified four-component model is closer than the original to the  MHD, both in having a deeper gap and also in having its pressure bump slightly weaker and further out.} 
\label{fig:full-model}
\end{figure}

\begin{figure*} 
\centering
\includegraphics[width=0.47\textwidth]{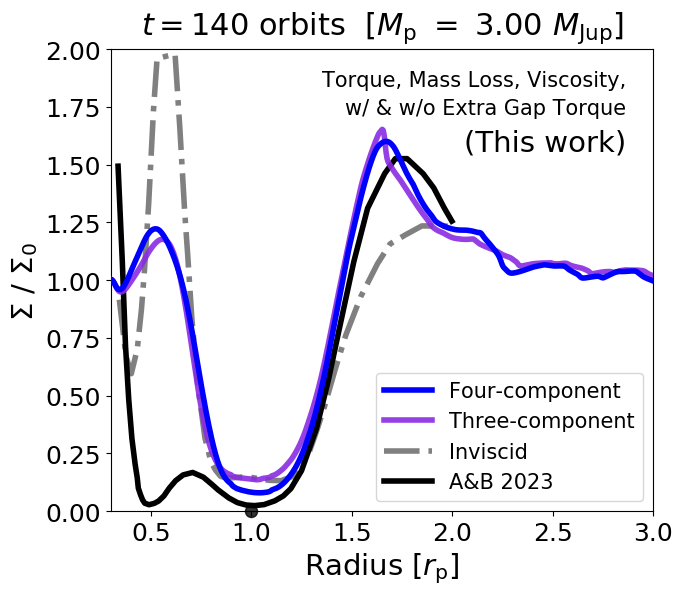}
\hspace{1.0em}
\includegraphics[width=0.47\textwidth]{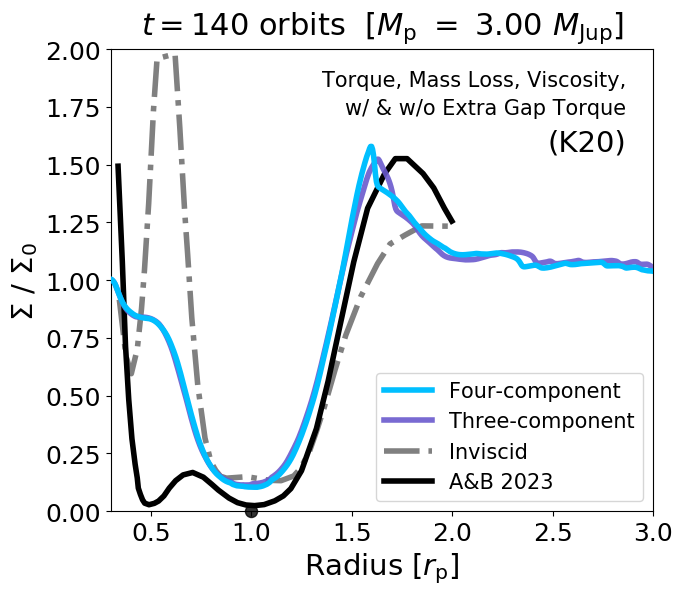}
\caption{Comparison of the three-component and four-component prescribed disc wind models from this work (\textit{left: blue and purple}) and by K20 (\textit{right: light blue and light purple}), the  MHD simulation (\textit{black}) by A\&B 2023, and their inviscid comparison run (\textit{grey dashed line}), all at $t = 140$. There is a noticeable difference in the gap depth with the model from this work, but no noticeable difference with the K20 model.} 
\label{fig:extra-torque}
\end{figure*}

To make the trough of the gap deeper, we added in extra torque in the gap as a fourth component. 
As Figure~\ref{fig:full-model} shows, the extra torque does indeed make the gap deeper with our model, although still not as deep as the MHD run. Unlike with our model, the extra torque does not actually make gap deeper with the K20 model, as highlighted in Figure~\ref{fig:extra-torque}. Another minor improvement is that the location of the pressure bump and general shape of the gap profile there are slightly better with our model than the K20 model.


\begin{figure*} 
\centering
\includegraphics[width=0.47\textwidth]{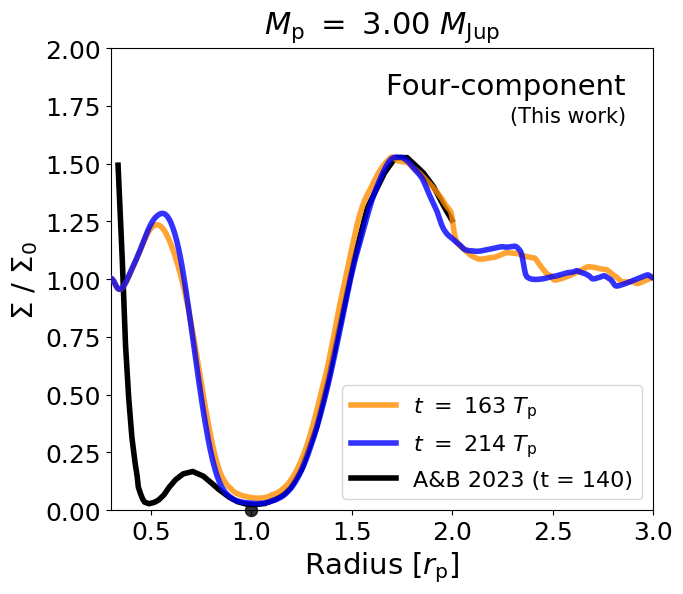}
\hspace{1.0em}
\includegraphics[width=0.47\textwidth]{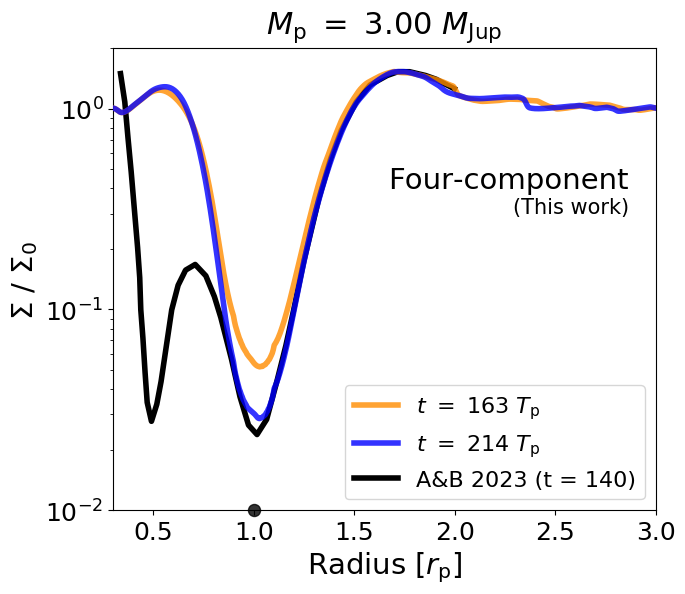}
\caption{Comparison of the four-component prescribed disc wind model from this work at $t = 163$ (\textit{orange}) and $t = 214~T_\mathrm{p}$ (\textit{blue}) to the  MHD simulation (\textit{black}) by A\&B 2023 at $t = 140~T_\mathrm{p}$. At $t = 163$, the outer gap edge matches perfectly. At $t = 214$, the entire outer region -- both the trough and the outer gap edge -- matches perfectly.} 
\label{fig:perfect-match}
\end{figure*}

Even though our four-component model already does well to match the MHD gap profile, we can get even better matches if we continue the simulation. As shown in Figure~\ref{fig:perfect-match}, the gap profile at the outer edge at $t = 163$ orbits just about perfectly matches the MHD simulation at $t = 140$ orbits, including the entire inner shoulder up to and past the peak of the pressure bump. The only part left to match is the trough. Once the gap reaches $t = 214$ orbits, the trough of the gap also just about matches the gap depth reached at the end of the MHD simulation. Meanwhile, the profile at the outer gap edge still aligns, creating a near-perfect match in the entire region exterior to the planet. This ``delay" of about 70 orbits to almost perfectly match the gap profile is consistent with the gap itself opening much slower, as we discuss in the next section. Overall, we conclude our four-component model can achieve a near-perfect match to the MHD gap profile, albeit at a noticeable delay.

\subsection{Gap profile evolution over time} \label{sec:gap-evolution}


\begin{figure} 
\centering
\includegraphics[width=0.45\textwidth]{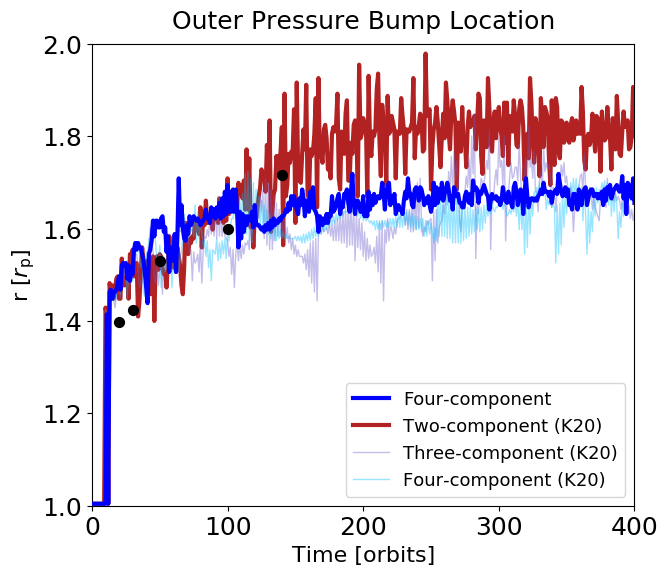} \\
\includegraphics[width=0.45\textwidth]{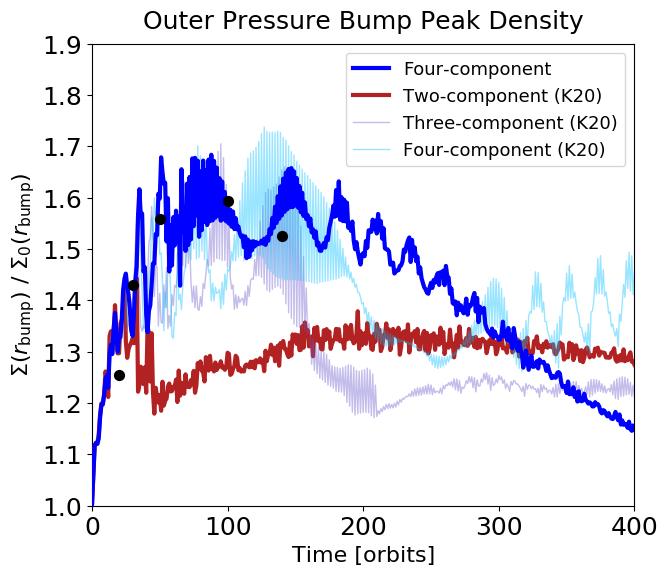}
\caption{Gap edge location (\textit{top}) and gap edge density (\textit{bottom}) over time with different numbers of components to the prescribed wind. The values from A\&B 2023 (\textit{black dots}) at five different times are shown for comparison. All of the models except the basic two-component model with no viscosity match the gap edge density and position reasonably well, indicating viscosity is the key component to match these features.} 
\label{fig:evolution}
\end{figure}

\begin{figure} 
\centering
\includegraphics[width=0.45\textwidth]{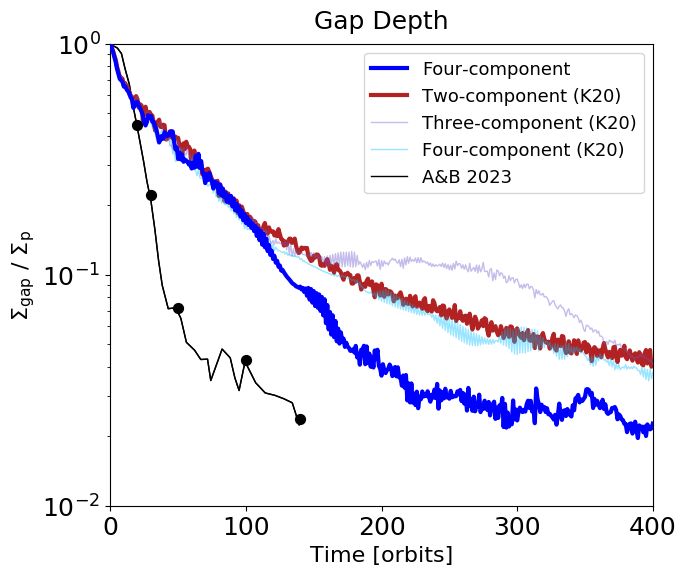}
\caption{Gap depth over time with different numbers of components to the prescribed wind. The values from A\&B 2023 (\textit{black line}), highlighted at five different times (\textit{black dots}), are shown for comparison. The full four-component model with the new torque profile is the only fit that approaches the depth reached with MHD, albeit with a time delay. \cbf {\textit{The value $\Sigma_\mathrm{gap}$ is defined to be the minimum azimuthally-averaged surface density in the gap within two radii of the planet's location.}}} 
\label{fig:evolution2}
\end{figure}

\begin{figure} 
\centering
\includegraphics[width=0.47\textwidth]{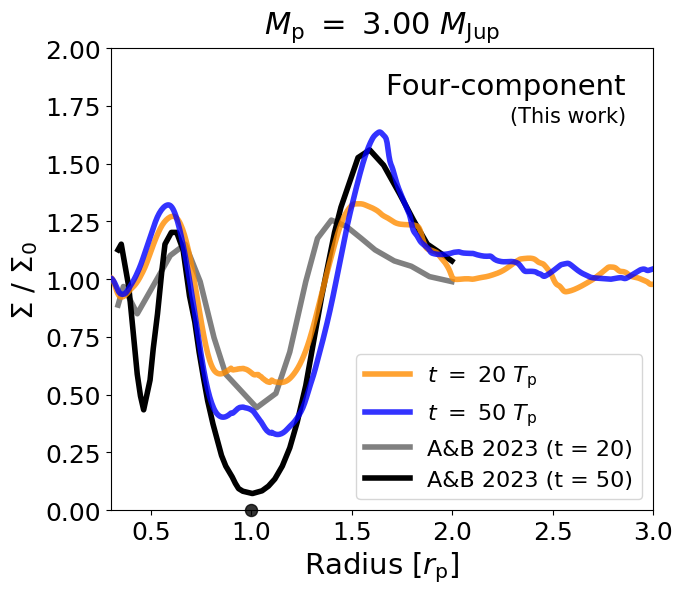}
\caption{Comparison of the four-component prescribed disc wind model from this work in the early stages at $t = 20$ (\textit{orange}) and $t = 50~T_\mathrm{p}$ (\textit{orange \& blue}) to the MHD simulation (\textit{grey \& black}) by A\&B 2023. The gap MHD simulation opens much faster and does not have the ``double gap" structure that appears with our prescribed model.} 
\label{fig:double-gap}
\end{figure}

Although we find a reasonably good fit to MHD gap profile near the end of the MHD simulations, the fits earlier on \cbf{in} the simulation were not as good. At the very beginning, the differences between our best-fit prescribed model and the MHD gap profiles were the largest. These differences lessen over time, \cbf{as illustrated in Figures~\ref{fig:evolution}~and~\ref{fig:evolution2}}. In particular, there are three main differences that the prescribed models do little to help address, each of which is noticeable in the gap profiles at early times, as depicted in Figure~\ref{fig:double-gap}.

First, the gap deepens much more quickly with MHD than with the prescribed models. Second, the MHD gap profiles are much more symmetric than the ones with the prescribed models. \cbf{Third, with MHD, the planet opens a gap directly at the location of the planet from the onset or close. In contrast, with the prescribed model, the planet much more clearly opens its gap from a shock length away, both interior and exterior to the planet. This creates a well-known  ``double gap" structure that is characteristic of planets in low-viscosity discs, particularly if the planet is lower mass.}
Each of these discrepancies \cbf{dwindles} over time as the gap gets deeper, but \cbf{is} still at least somewhat noticeable by the comparison point at the end of the MHD simulations.

The double gap, as well as the speed at which the gap deepens, both relate to how the gap depletes. In our simulations, some of the gas remains trapped at the L5 Lagrange point $60^{\circ}$ co-orbital behind the planet. Gas collects here because a vortex-like structure develops at the L5 point, a well-known phenomenon in hydrodynamic simulations with a planet \citep[e.g.][]{montesinos20}. This gas at L5 both creates the double gap and slows the depletion of the gap.
In contrast, the extra torque caused by the wind in the MHD simulation helps clear out all of the material in the gap, even at the L5 point. \mklrc{isn't this extra torque accounted for in the prescribed model by boosting the gap torque?} Even with the extra torque in our prescribed model, we do not see that happen in our simulations. As a result, the MHD gap profiles deepen much quicker than ours, especially at the onset. 

That quickness is somewhat surprising given that the density in the gap isn't that much lower at the onset. We had expected that the gap would not deplete faster until much later because the extra torque in the gap is expected to arise from the gap's much lower density.
This early discrepancy is why we \cbf{use} such \cbf{a} high $K_\mathrm{gap} = 5$ level of extra torque in the gap and do not use any tapering or density dependence near the beginning. The level recommended by A\&B 2023 was $K_\mathrm{gap} = 3$ to 5, but even with the high level we use, our gaps still lag behind, most noticeably in the gap depth. That lagging may suggest the deeper gaps are not just due to the extra torque in the gap; otherwise \cbf{they} shouldn't arise so early on.

\cbf{Beyond the time when we get a near-perfect match at $t = 214$ orbits, the rate at which the gap deepens begins to slow down. Over the next hundred to two hundred orbits, the gap depth does not significantly increase, staying between about 1/32 to 1/42. During this range of times, we do not know what to expect from A\&B 2023 since their simulations do not extend this far. In \cite{wafflard23} though, they ran some of their simulations a little more than twice as long and find that the gap depth does eventually reach a steady state after the initial few hundred orbits. As such, it may bode well for our model that the fast rate at which the gap deepens does not continue indefinitely.}





\subsection{Different planet masses} \label{sec:other-planet-mass}

\begin{figure*} 
\centering
\includegraphics[width=0.47\textwidth]{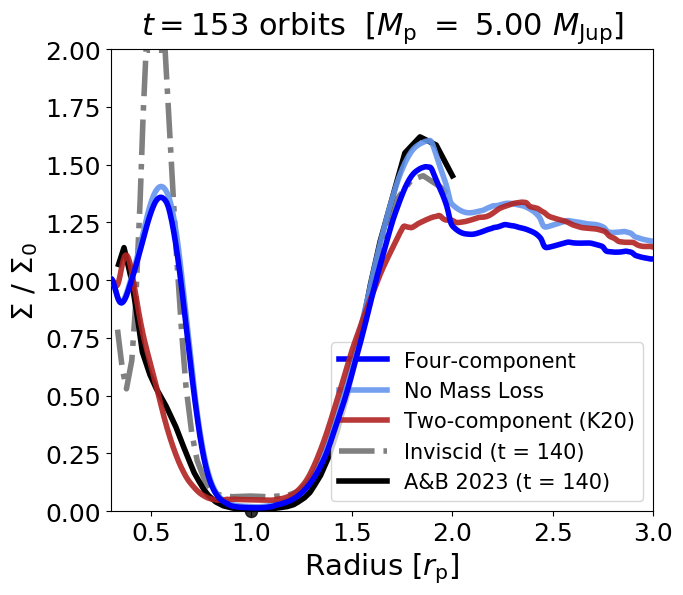}
\hspace{1.0em}
\includegraphics[width=0.47\textwidth]{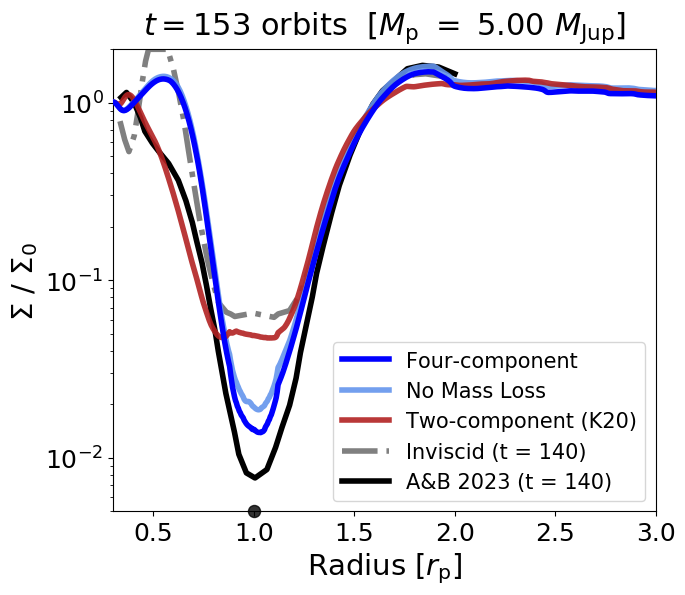}
\caption{High-planet-mass comparison of the modified four-component (MHD-based torque, MHD-based mass loss, viscosity \& extra gap torque) prescribed disc wind model (\textit{blue}) to the same model except without mass (\textit{cornflower blue}), and the K20 model (\textit{red}) all at $t = 153$, as well as the MHD simulation (\textit{black}) by A\&B 2023 and their inviscid comparison run (\textit{grey dashed line}) at $t = 140$. The planet mass is $5~M_\mathrm{Jup}$. Like at intermediate planet mass, the viscosity elevates the pressure bump at the outer gap edge. Unlike at intermediate planet mass, the case without mass loss gives a better fit while the case with mass loss better resembles the inviscid run.} 
\label{fig:high-mass}
\end{figure*}

\begin{figure} 
\centering
\includegraphics[width=0.47\textwidth]{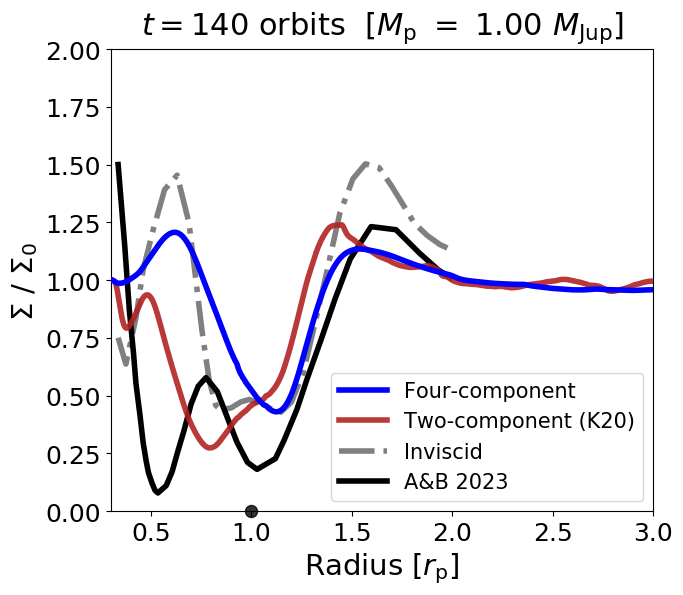}
\caption{Low-planet-mass comparison of the modified four-component (MHD-based torque, MHD-based mass loss, viscosity \& extra gap torque) prescribed disc wind model (\textit{blue}) to the K20 model (\textit{red}), as well as the MHD simulation (\textit{black}) by A\&B 2023 and their inviscid comparison run (\textit{grey dashed line}) at $t = 140$. The planet mass is $1~M_\mathrm{Jup}$. Unlike at intermediate planet mass, viscosity does not elevate the pressure bump at the outer gap edge and as a result does not help produce a better fit.} 
\label{fig:low-mass}
\end{figure}

Like our fiducial test case, we can reproduce the gap profiles for the 5 Jupiter-mass (5 thermal-mass) planet case, the highest-mass planet simulated by A\&B 2023 reasonably well with the same model parameter values. \mklrc{with the same model parameter values?} Figure~\ref{fig:high-mass} shows how the gap profiles match up. Without any prescribed wind or viscosity, the gap edge shoulder already follows the MHD reasonably well, but \cbf{there is still a big discrepancy in that} the gap itself is way too shallow. Like with a 3 Jupiter-mass planet, the viscosity similarly helps strengthen the pressure bump at outer gap edge, while the extra gap torque helps deepen the gap. One small difference is that the outer gap edge has the correct amplitude without mass loss. Adding in mass loss drops the amplitude a bit so that the full model better aligns with the inviscid case near the gap edge \cbf{instead of the MHD we are trying to match}. The main strengths are that the shoulder and gap edge location align with the MHD gap profile almost perfectly, just like in the fiducial case.


On the other hand, the gap profiles for the lowest-mass case with a 1 Jupiter-mass (1 thermal-mass) planet are much more difficult to reproduce. Figure~\ref{fig:low-mass} compares the gap profile for this case to the corresponding viscous and inviscid comparison runs by A\&B 2023. 
Unlike in the higher-mass planet cases, the inviscid pressure bump at the outer gap edge \cbf{actually has a} higher amplitude than the MHD, not lower. The pressure bump does not end up as weak because the planet is too low-mass to push the pressure bump away from it. With the pressure bump already too strong, adding in viscosity does not help the way it does in the higher-mass planet cases. It does weaken the pressure bump like the MHD \cbf{case with a 1 Jupiter-mass planet}; however, the profiles still generally do not come close to matching, mainly because of the trough. The gap depth is underestimated by such a large amount that it does not matter if the viscosity helps improve \cbf{the} profile at the outer gap edge. With the structure in the gap itself being so different, the gap profile as a whole ends up with a fundamentally different structure. 

We see that issue of an underestimated gap depth messing up the entire gap profile in the higher-mass planet cases as well, but only at earlier times. In those cases, however, the planet is massive enough that the gap eventually becomes much deeper later on. As a result, the gap depth in our simulations and the MHD simulations eventually converge to about the same level, even if the percentage difference between the two is still large. With a lower-mass planet, the gap is much shallower and thus, the large percentage difference between the two has a \cbf{more} significant effect on the rest of the gap profile.


Overall, we can reproduce the high-mass planet gap profiles reasonably well, but cannot match the lower-mass case. We find it promising that we can reproduce the gap profiles with the same model parameters for both the 3 $M_\mathrm{Jup}$ and 5 $M_\mathrm{Jup}$ cases even though there are moderate differences in the MHD runs between their measured profiles for $\alpha$, torque, and mass loss. 

We do not think it is surprising that the low-mass case is more difficult to match because the planet is only at the thermal mass. We interpret the feasibility of which gap profiles we can match as evidence that the more massive planets play a more dominant role in shaping the gap structure. When the planet is lower mass and barely able to open a gap, the wind plays a more integral role in shaping the gap structure, making it more difficult to capture the MHD effects associated with the wind with just a prescription.


\subsection{Role of vortices} \label{sec:vortices}

\begin{figure} 
\centering
\includegraphics[width=0.48\textwidth]{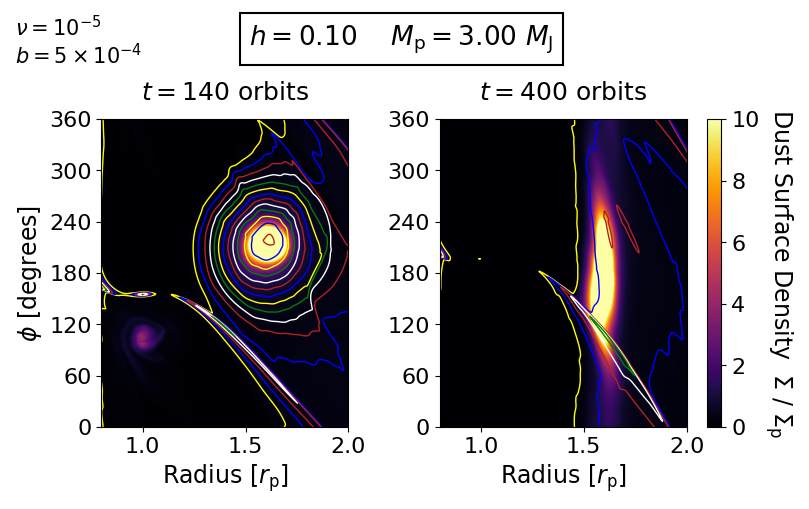} \\
\vspace*{0.5em}
\hspace*{0.5em}
\includegraphics[width=0.48\textwidth]{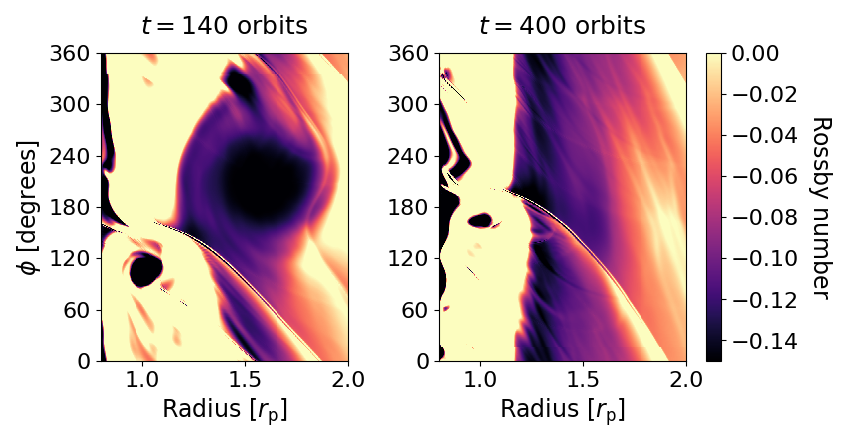}
\caption{Snapshots showing the evolution of the the dust density (\textit{top panels}) and Rossby number (\textit{bottom panels}) in the presence of a super-thermal mass planet ($M_\mathrm{p} = 3.0~M_\mathrm{Jup}$) in a disc with our fiducial full wind model ($b = 5 \times 10^{-4}$ and $\nu = 10^{-5}$). Gas density contours (at $\Sigma / \Sigma_\mathrm{p} = 0.4, 0.5, 0.6,$ etc.) are overlaid. The vortex is compact but quickly fades before 500 orbits due to the high viscosity.}
\label{fig:vortex-Mt3}
\end{figure}

\begin{figure} 
\centering
\includegraphics[width=0.48\textwidth]{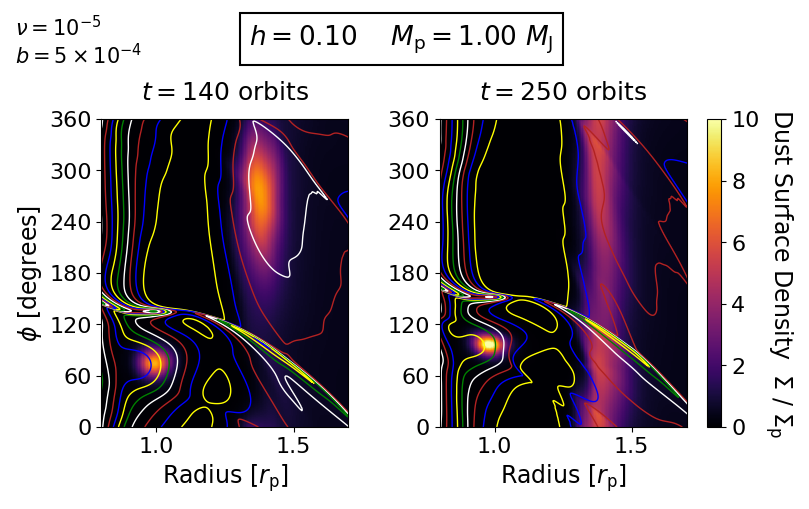} \\
\vspace*{0.5em}
\hspace*{0.5em}
\includegraphics[width=0.48\textwidth]{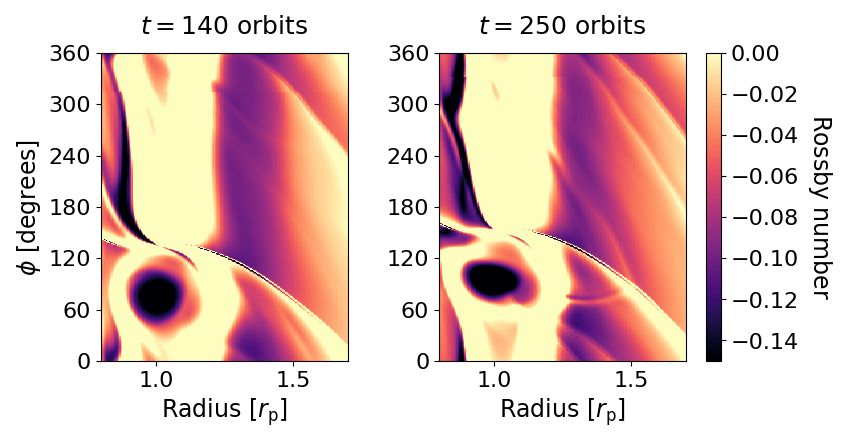}
\caption{Snapshots showing the evolution of the the dust density (\textit{top panels}) and Rossby number (\textit{bottom panels}) in the presence of a thermal mass planet ($M_\mathrm{p} = 1.0~M_\mathrm{Jup}$) in a disc with our fiducial full wind model ($b = 5 \times 10^{-4}$ and $\nu = 10^{-5}$). Gas density contours (at $\Sigma / \Sigma_\mathrm{p} = 0.4, 0.5, 0.6,$ etc.) are overlaid. The vortex is elongated and spreads into a ring by around 250 orbits in both the gas and dust.} 
\label{fig:vortex-Mt1}
\end{figure}

Our main interest in prescribed disc winds was to see if they can match the 1D MHD planetary gap profiles, but the gap profiles themselves are not truly 1D due to vortex formation at the gap edge. 

In general, vortices can arise at planetary gap edges \citep{li05, hammer17, hammer21, hammer23} through the Rossby Wave instability if the disc has sufficiently low viscosity \citep{deValBorro07}. The specific criteria that makes the disc unstable to the RWI is an inverse vortensity maximum becoming too sharp \citep{mkl12a, ono16}, where the inverse vortensity is surface density $\Sigma$ divided by vorticity $\omega = (\vec \nabla \times \vec{v})_\mathrm{z}$.\mklrc{do you need to define the inverse vortensity?} That criteria may be satisfied at the maximum located nearly coincident with the pressure bump at a gap edge. However, higher viscosities can smooth out the gap edge enough to prevent the RWI from occurring. The resulting vortices can be compact or elongated, which largely depends on whether the local Rossby number $\mathrm{Ro} = (\vec \nabla \times (\vec v - \vec v_\mathrm{K}))_\mathrm{z} / 2 \Omega$  is above or below the critical value of $-0.15$ respectively \citep{surville15, hammer21}.

\mklrc{this section sees like a discussion, not results}

Like in the MHD simulations by A\&B 2023, all of the planets in our fiducial simulations trigger vortices. Also like their work, the two higher-mass planets trigger stronger vortices, as depicted in Figure~\ref{fig:vortex-Mt3}, while the lowest-mass planet triggers a noticeably weaker vortex, as depicted in Figure~\ref{fig:vortex-Mt1}. The stronger vortices are compact with Ro $< -0.15$, while the weaker vortex is elongated. \mklrc{need to define rossby number} The stronger vortices last about 400 to 500 orbits in both the gas and dust, while the weaker vortex lasts about half as long at around 200 orbits. All of the vortex lifetimes are relatively short because of the relatively high viscosity.

As the MHD simulations have vortices, it would be ideal if our simulations had vortices with similar morphologies in order to match the gap profiles. Vortices can affect the gap profiles in a few ways. They can transport material outwards from along the outer gap edge shoulder into the vortex itself when the instability is triggered. If they are compact, they can also drive accretion as their spiral waves transport angular momentum.

Allowing vortices to occur by not having too high of a viscosity indeed helps us match the profiles for the highest-mass planets, but not the lower-mass one. We believe the lower-mass one still does not match for two main reasons. First, the trough of the gap is way too shallow compared to the MHD and that has little to do with the vortex. Second, the vortex itself is elongated and with this weaker structure, it does not have as much effect on the gap profile. 




\subsection{Role of inner MHD gap} \label{sec:inner-gap}


\mklrc{consider skipping this model. is the delay a significant enough caveat to fix? one could argue that we're more likely observe steady states. for this section to be useful, one needs to present the detailed implementation so people can follow it. but it seems difficult to prescribe MHD gaps because its formation is through an instability} 

\begin{figure} 
\centering
\includegraphics[width=0.47\textwidth]{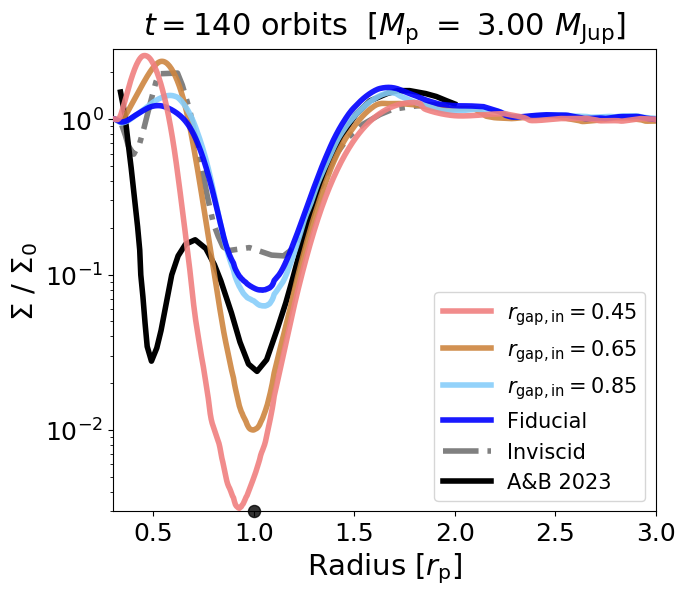} \\
\caption{Gap depth comparison with different-sized "gaps" for where the extra gap torque is applied. The outer edge of the gap is normal, while the inner edge is moved inwards to $r_\mathrm{gap,in}$ as a crude way to resemble the MHD gap from A\&B 2023. With more gas cleared out interior to the planet, the entire planetary gap becomes much deeper. As such, the lack of an MHD gap in our fiducial case may explain part of why the gap is shallower.}
\label{fig:MHD-gap}
\end{figure}

One of the main differences between our prescribed disc wind model and the disc wind simulations by A\&B 2023 is the secondary MHD-driven gap located interior to the planet. We ignored this feature entirely in our fiducial simulations and were still able to reproduce the rest of the gap structure away from this feature, albeit at a time delay, particularly for the depth of the gap. As such, we were interested in investigating whether the MHD gap could be partially responsible for the time delay in the opening of the gap.

As we were unable to prescribe an inner MHD gap, we instead tested what would happen if we expanded the planetary gap region to include more of the interior region of the disc. Specifically, rather than just apply the extra gap torque at a level $K_\mathrm{gap} = 5$ in the usual region, we \cbf{moved the inner boundary of this region further in and also} applied the extra torque in the inner part of the disc over a range $r_\mathrm{gap,in} \le r \le r_\mathrm{p}$, leaving out smoothing at the inner edge. We tested three values of $r_\mathrm{gap,in} = 0.45$, 0.65, and 0.85 $r_\mathrm{p}$. This setup mimics the later stages of the MHD simulations where the planetary gap and inner MHD gap have some overlap. It may not capture what happens in the early stages of the MHD simulations before there is any overlap. As Figure~\ref{fig:MHD-gap} shows, the planetary gap does indeed open up much faster when it encompasses more of the inner disc, supporting the notion that the MHD gap contributes to the gap deepening at a faster rate.



We note that although the inner MHD gap appears to help deepen the gap, it is not the sole reason for the gap opening more slowly. As we had already pointed out, the lack of depletion of the gas at the Lagrange point also contributes to the slower gap-opening process, particularly at earlier times before it depletes.


\subsection{Alterations to best-fit model} \label{sec:alterations}

\begin{figure*} 
\centering
\includegraphics[width=0.47\textwidth]{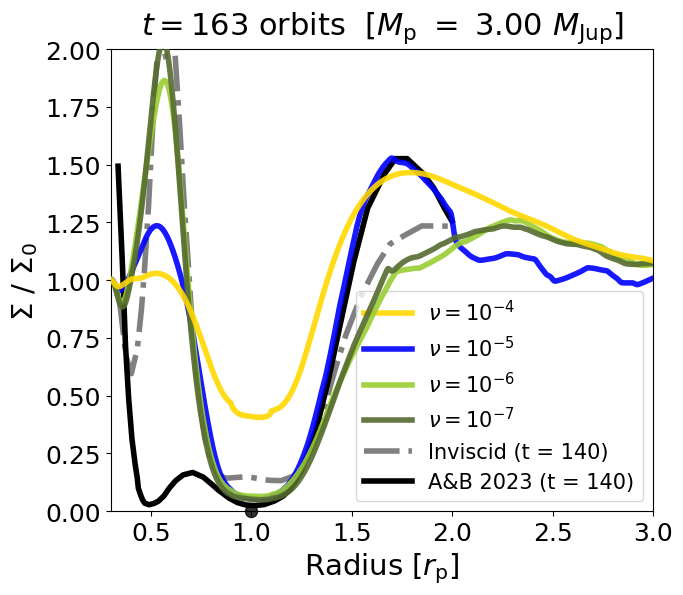}
\hspace{1.0em}
\includegraphics[width=0.47\textwidth]{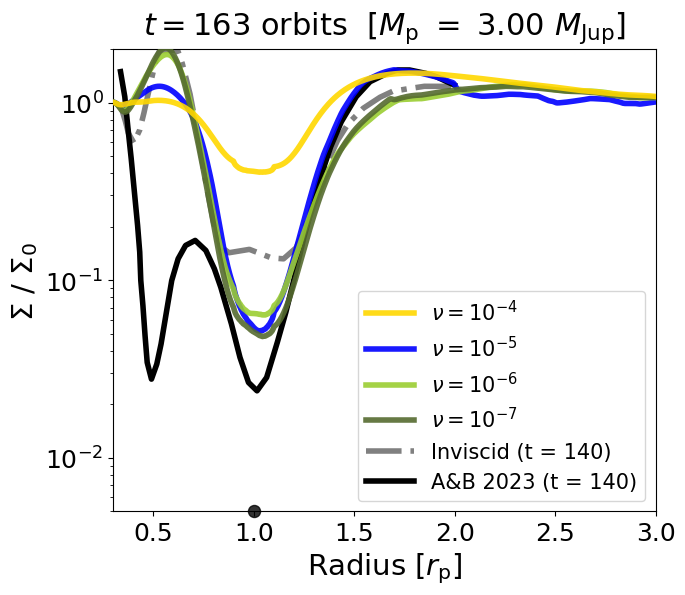}
\caption{Comparison of the modified four-component (MHD-based torque, MHD-based mass loss, viscosity \& extra gap torque) prescribed disc wind model (\textit{blue}) to the same model with different viscosities all at $t = 163$, as well as the MHD simulation (\textit{black}) by A\&B 2023 and their inviscid comparison run (\textit{grey dashed line}) at $t = 140$. With higher viscosity, the amplitude of the pressure bump increases while the amplitude of the gap depth decreases (although the latter effect is not as strong until later times).} 
\label{fig:viscosity}
\end{figure*}

\begin{figure} 
\centering
\includegraphics[width=0.47\textwidth]{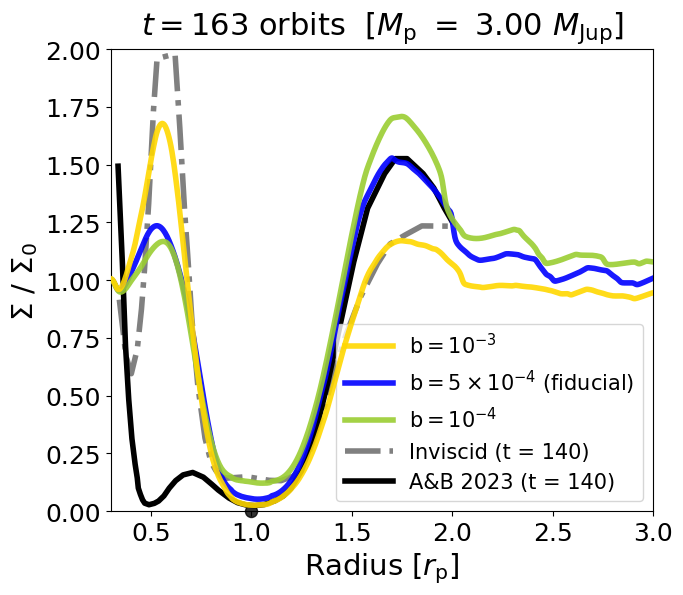}
\caption{Comparison of the modified four-component (MHD-based torque, MHD-based mass loss, viscosity \& extra gap torque) prescribed disc wind model (\textit{blue}) to the same model with different wind strengths all at $t = 163$, as well as the MHD simulation (\textit{black}) by A\&B 2023 and their inviscid comparison run (\textit{grey dashed line}) at $t = 140$. With stronger winds, the amplitude of the pressure bump drops and the amplitude of the gap depth increases.} 
\label{fig:wind-strengths}
\end{figure}

Our best-fit model uses the wind strength $b = 5 \times 10^{-4}$ taken from the simulations by A\&B 2023 and a viscosity $\nu = 10^{-5}$ approximately one-sixth the level from their work. To show these values produce the best fit, we also tested other higher and lower values for both parameters.

With different viscosities, we observe three expected regimes, as shown in Figure~\ref{fig:viscosity}. At high levels, the viscosity dominates over the wind and is the main factor determining the shape of the gap profile aside from the mass loss. The high viscosity also suppresses vortices, resulting in a fundamentally different gap structure where the gap is more much shallow and the outer gap edge has a different shape. At low viscosity, the wind dominates over the viscosity, leaving the gap profile mainly determined by the wind and the vortices that exist. The gap is deeper and the outer pressure bump is further from the planet. Our fiducial model lies in-between at intermediate viscosity, where the wind-driven accretion rate is higher than the viscosity-driven accretion rate (see Section~\ref{sec:viscosity})\mklrc{how do we compare these two? by comparing the vr induced by viscosity and vr induced by wind? TO DO: add this comparison in.}, but the viscosity is still strong enough to matter. The gap depth reaches an intermediate value, while the amplitude of the outer gap edge is actually stronger than with higher or lower viscosity.

With different wind strengths, we see a smoother transition in the gap profiles from higher to lower values, as shown in Figure~\ref{fig:wind-strengths}. Stronger winds create a much weaker pressure bump and a much deeper gap. At $t = 163$, the stronger wind already has the correct gap depth, in contrast to our fiducial model.
We still prefer our fiducial model, however, because we believe the L5 gas and the lack of an inner MHD gap are the reasons the gap depth is underestimated.


\section{Extensions} \label{sec:extensions}

Now that we have tested our prescribed disc model against the MHD simulations on which it is based, we apply it to a wider parameter space and longer simulations. We test the model's applicability while also exploring \cbf{a disc wind's} potential effects in regimes beyond what MHD simulations have explored so far. These tests require minor modifications to the setup that are described directly in the relevant subsection.


\subsection{Varying disc parameters} \label{sec:disc-parameters}

\mklrc{***MKL editing bookmark***}

\begin{figure} 
\centering
\includegraphics[width=0.47\textwidth]{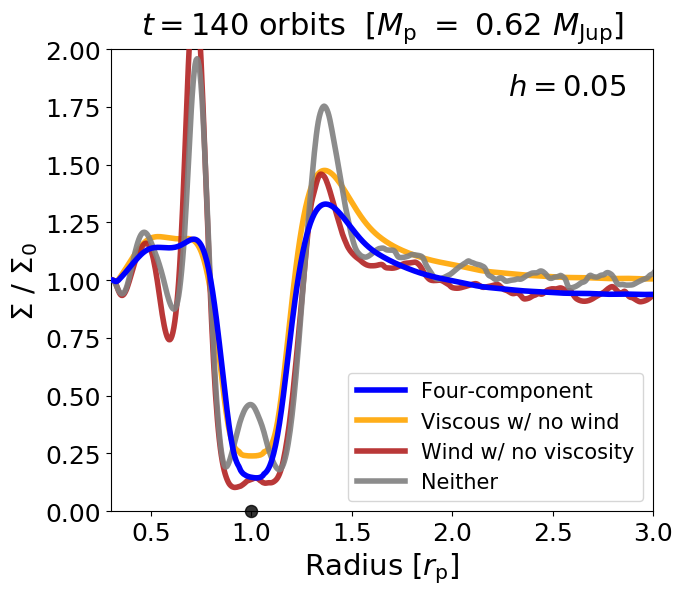} \\
\includegraphics[width=0.47\textwidth]{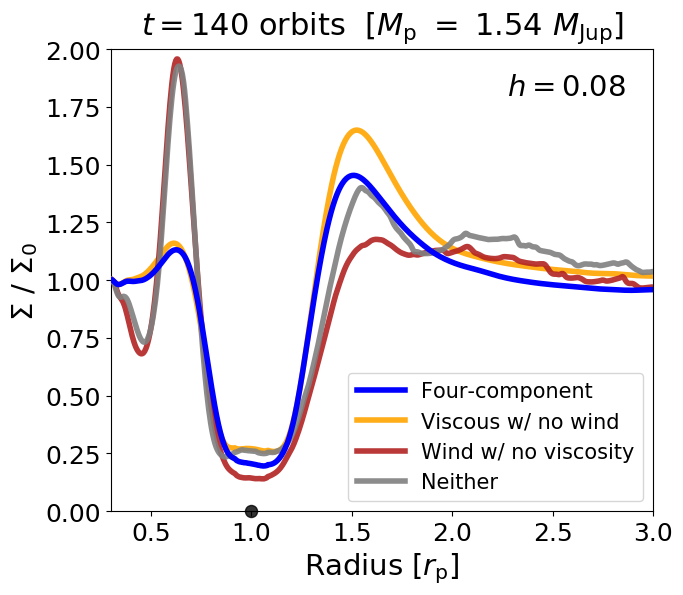}
\caption{Low-aspect ratio comparison of the the full model (MHD-based torque, MHD-based mass loss, viscosity \& extra gap torque) compared to just viscosity (\textit{gold}), just wind (\textit{red}), and neither (\textit{grey}). The planet mass is $1~M_\mathrm{Jup}$. Unlike at the fiducial aspect ratio of $h = 0.1$ from A\&B 2023, viscosity does not elevate the pressure bump at the outer gap edge. At  $h = 0.05$, the pressure bump drops with viscosity. At  $h = 0.08$, the pressure bump is not significantly different until after $t = 140$. As a result, we are not optimistic viscosity would help produce a better fit, particularly at lower aspect ratios.}
\label{fig:h05}
\end{figure}

With the new prescribed model capable of reproducing gap profiles for high-mass planets and a relatively thick disc aspect ratio of $h = 0.1$, we now test whether it can do the same at thinner aspect ratios of $h = 0.05$ and 0.08. \cbf{With $h = 0.08$, we still focus on a 3 thermal-mass planet, corresponding to $M_\mathrm{p} = 1.54~M_\mathrm{Jup}$. With $h = 0.05$, we instead focus on a 5 thermal-mass planet, corresponding to $M_\mathrm{p} = 0.62~M_\mathrm{Jup}$} For each of the two aspect ratios, we test four cases:
\begin{enumerate}
  \item our full prescribed model,
  \item a control case with just our fiducial viscosity,
  \item a control case with just the wind and no viscosity, and
  \item a control case with neither the wind or viscosity.
\end{enumerate}
Although we do not have MHD simulations to compare to for these thinner discs, we can still compare the prescribed wind model to wind-less viscous and inviscid counterparts to see if the same trends occur.

For the thinnest disc at $h = 0.05$, we find that the trend between the different models better resembles the low-mass $1~M_\mathrm{Jup}$ case that we could not reproduce. In particular, the inviscid simulation produces the strongest pressure bump, stronger than either the viscous case or the prescribed model, as illustrated in the top panel of Figure~\ref{fig:h05}. As such, adding viscosity to the prescribed model no longer helps strengthen the inviscid pressure bump because it is already too strong.

We believe our prescribed model fares slightly better with the intermediate aspect ratio of $h = 0.08$. Although the viscosity still does not help elevate the pressure bump, at the very least the pressure bump does not drop either. Instead, the pressure bump location and amplitude are about the same with our full model as in the inviscid control case with no wind, which is illustrated in the bottom panel of Figure~\ref{fig:h05}. The gap edge shoulder is one part that is different, owing to the gap being deeper in the case with our full model. Overall, the trends for this set of cases is a bit different from the $1~M_\mathrm{Jup}$ case, although it still does not follow the trend from the 3 or $5~M_\mathrm{Jup}$ cases that we could reproduce.

Since we could not reproduce the $1~M_\mathrm{Jup}$ case and the only cases we could reproduce are the ones where the viscosity elevates the outer gap edge pressure bump, we are less optimistic about the ability of our model to reproduce gap profiles at lower aspect ratios. It may still be possible for our model to work, though, since we expect \cbf{it might be possible to reproduce the $1~M_\mathrm{Jup}$ gap profile if the trough were deeper.}
As such, for our model to have success at lower aspect ratios, we think that success would be more likely with more massive planets that open gaps deep enough for the trough discrepancy not to matter.

\mklrc{do you expect longer simulations for lower Mp and h cases would help produce deeper gaps? A: It should produce deeper gaps, but I do not know if that will help because for the 1~M_p case at h=0.1, I'm not optimistic that the prescription will ever help.}


\subsection{Vortex trigger criteria} \label{sec:criteria}

\begin{figure} 
\centering
\includegraphics[width=0.48\textwidth]{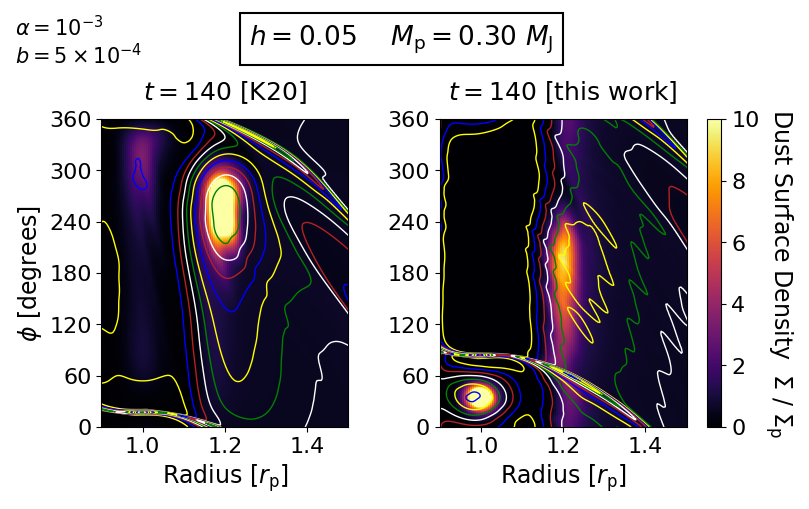} \\
\vspace*{0.5em}
\includegraphics[width=0.48\textwidth]{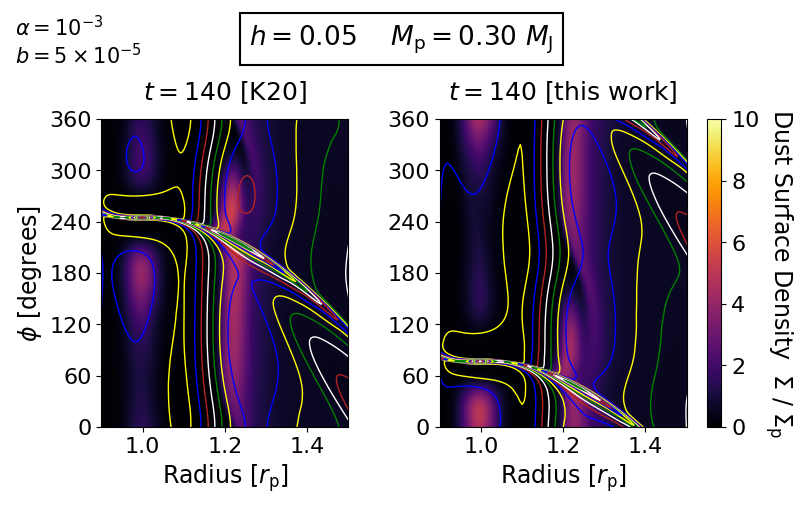}
\caption{Dust density comparison for vortices with the prescribed wind models from K20 (K20: \textit{left panels}) and this work (\textit{right panels}). Gas density contours (at $\Sigma / \Sigma_\mathrm{p} = 0.4, 0.5, 0.6,$ etc.) are overlaid. A strong wind (\textit{top}) makes the vortex stronger with both models compared to a weaker wind (\textit{bottom}), but the effect is much more pronounced with the K20 model.} 
\label{fig:trigger}
\end{figure}

For planets to trigger vortices at their gap edges when they open up gaps, the disc must have a sufficiently low viscosity. It has been suggested that magnetic disc winds can help alleviate this restriction and allow stronger vortices to form at higher viscosities than usual \citep{wu23}. As that was done with a variant of the K20 model first introduced by \cite{elbakyan22}, we test if the same result can be obtained with our new prescribed wind model\cbf{, but find that it only occurs with the K20 model}.

\subsubsection{Criteria with $h = 0.05$} \label{sec:h05-vortices}

To perform this test, we ran additional simulations with a Saturn-mass planet $M_\mathrm{p} = 0.3~M_\mathrm{Jup}$ and a disc with a thinner aspect ratio $h = 0.05$, a surface density power law of $p = 1$, and $\alpha = 10^{-3}$
instead of the usual constant viscosity. This $\alpha$ corresponds to $\nu = 2.5 \times 10^{-6}$ at the location of the planet. \mklrc{relation between nu and alpha needs to be stated somewhere} We focus on two wind strengths, our fiducial value of $b = 5 \times 10^{-4}$ and a weaker value of $b = 5 \times 10^{-5}$, but we also tested other wind strengths from $10^{-6} \le b \le 10^{-3}$. One set of simulations uses the K20 model including viscosity and the other set uses our new model. With no wind or low wind ($b \le 5 \times 10^{-5}$) prescribed through either our model or the K20 model, the planet only triggers a weak elongated vortex that decays within 100 orbits after it formed. As expected from \cite{wu23}, with a strong wind ($b \ge 5 \times 10^{-4}$) and the K20 model, the vortex is indeed stronger, as shown in the top left panel of Figure~\ref{fig:trigger}. It has a compact structure and shows little sign of decay after 1000 orbits.

With our new model, though, the vortex does not get stronger with stronger winds, as shown in the right panels of Figure~\ref{fig:trigger}. Regardless of the wind strength, it decays within 100 orbits. To determine how it can be so different, we compared the vortensity profiles, the criteria for triggering the RWI, with the different wind models. Although there were slight differences, we did not notice any obvious change that would indicate the other model would produce a stronger vortex.

\subsubsection{Criteria with fiducial $h = 0.10$} \label{sec:h10-vortices}

\cbf{Lastly, we checked if a similar pattern of vortex strengths would occur for our fiducial aspect ratio $h = 0.1$. With a planet of the same number of thermal masses ($M_\mathrm{p} = 2.4~M_\mathrm{Jup}$) grown over 300 orbits following the standard FARGO3D $\sin^2$ tapering, we likewise find that the K20 model results in a stronger longer-lived vortex with $b = 5 \times 10^{-4}$, while our model does not. Similarly, neither model results in a long-lived vortex with a weaker $b = 5 \times 10^{-5}$ wind. This same pattern of vortex strengths as at h = 0.05 shows it does not just occur at low aspect ratios.}

\mklrc{any speculation as to why the new wind model does not promote vortex formation? W23's argument was that wind-driven accretion helps to push material towards the planet, thereby increasing the bump. but do not we make the same argument with viscosity-driven accretion promoting the bump? A: I do not think that reasoning is so robust, because it does not obviously increase the bump.}


\subsection{Long-term evolution} \label{sec:longterm}


\begin{figure} 
\centering
\includegraphics[width=0.45\textwidth]{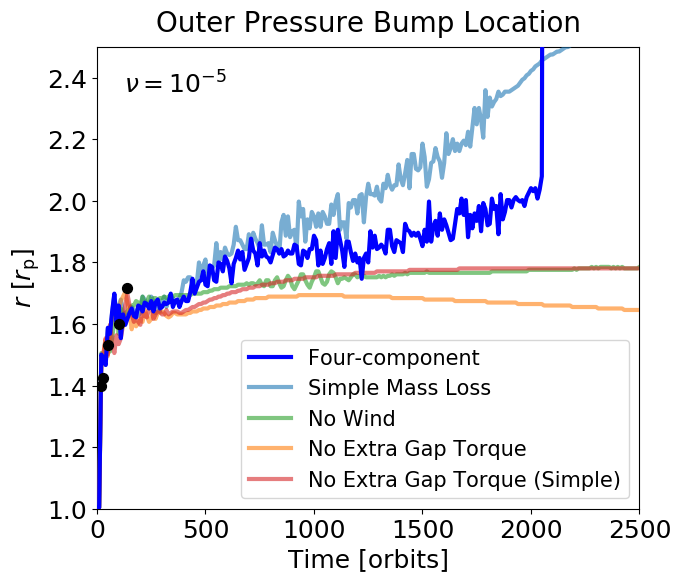}
\includegraphics[width=0.45\textwidth]{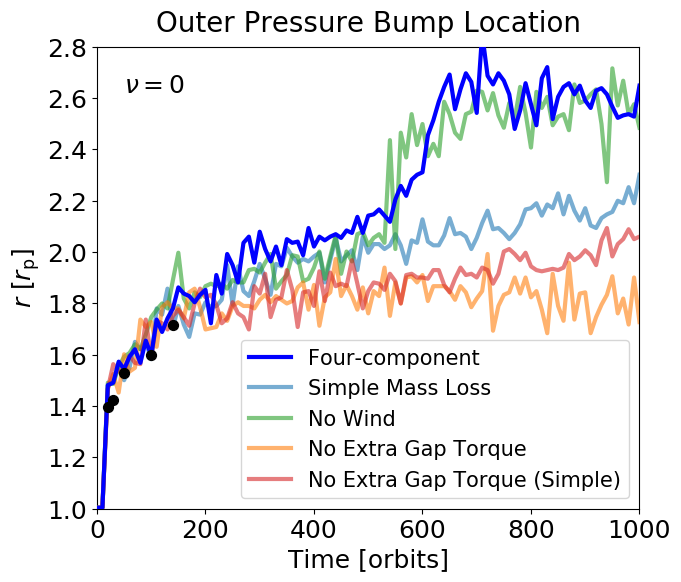}
\caption{Pressure bump location with viscosity (\textit{top}) and no viscosity (\textit{bottom}) over time with three different mass loss prescriptions, and with no wind. The values from A\&B 2023 (\textit{black dots}) at five different times are shown for comparison. \cbf{With and without viscosity, removing the extra gap torque causes the pressure bump to move inwards compared to the case with no wind. With viscosity and the extra gap torque, the pressure bump instead moves further outwards with either mass loss prescription.}}
\label{fig:long-term}
\end{figure}


\begin{figure} 
\centering
\includegraphics[width=0.47\textwidth]{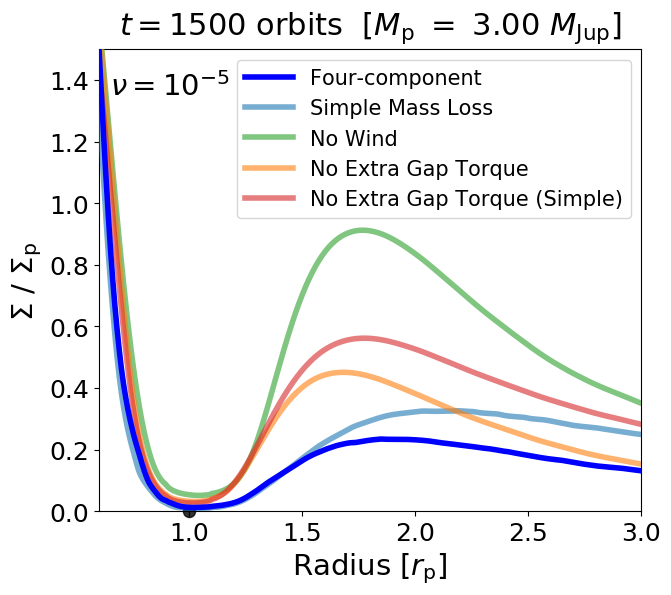}
\caption{Comparison of the gap profiles at $t = 1500~T_\mathrm{p}$ with viscosity between different mass loss prescriptions, with and without the extra gap torque, and with no wind, focusing on elucidating the pressure bump locations presented in Figure~\ref{fig:long-term}. The cases with no extra gap torque have the innermost pressure bumps. The cases with extra gap torque have the outermost pressure bumps. The pressure bump in the case with no wind lies in-between.} 
\label{fig:late-gap}
\end{figure}

One of the main drawbacks of MHD simulations of planets with disc winds thus far is the length of the simulations. Because of high computational expense, the simulations in A\&B 2023 were only run for 140 planet orbits, which is insufficient \cbf{for the gap profiles to reach a steady state}. Other simulations by \cite{wafflard23} and \cite{hu25} were likewise only run for a few hundred orbits. As such, one of our main interests in developing a prescribed disc wind model 
was to enable simulations to run for thousands of orbits, enough time to reach more of a steady state. \mklrc{this motivation should also be in the big intro. A: Added!}

We first tested running additional simulations with our fiducial parameters out to $3000~T_\mathrm{p}$. Because the location of the outer gap edge continues to move outwards more than in our shorter simulations, we moved the outer edge of the domain further away from the planet from $r = 4$ to $r = 5.85$. We also increased the radial resolution from 768 cells to 1152 cells to keep the resolution the same. We tested two sets of cases, one set with our fiducial viscosity and one with no viscosity, and mainly varied the mass loss prescriptions \cbf{and the extra gap torque. Each set consists of 
\begin{enumerate}
  \item our full prescribed model,
  \item our full model but with the simpler mass loss prescription from K20,
  \item our full model but with no extra gap torque,
  \item our full model but with no extra gap torque and with the simpler mass loss prescription from K20, and
  \item a control case with no wind.
\end{enumerate}}
Every setup featured our fiducial $3~M_\mathrm{Jup}$ planet and $ h = 0.1$ disc, while the viscous cases all had $\nu = 10^{-5}$.

As the time evolution in Figure~\ref{fig:long-term} shows, we find that the pressure bump moves much further away from the planet when there is a wind compared with no wind. \cbf{The extra gap torque is the primary factor moving the pressure bump away, particularly in the viscous cases. With the higher accretion rate induced by the extra torque, the gap is constantly being filled in by the outer disc, depleting the disc just outside the gap. As that inner part of the outer pressure bump is depleted, the peak in the pressure bump moves outwards, as illustrated in the gap profiles in Figure~\ref{fig:late-gap}.} 

\cbf{This outward motion of the outer gap edge is analogous to the outward motion of cavity edges found by \cite{martel22} in their study of wind-driven cavities, which was likewise caused by a sharp change in the accretion rate near that edge. When we remove the extra gap torque, the pressure bump instead ends up even closer to the planet than in the case with no wind, as had been found in previous studies of prescribed disc winds.}

\cbf{With no viscosity, the pressure bump ends up much further away from the planet, both in our fiducial case and with no wind. In this set of cases, the pressure bump can also end up closer to the planet with just the simple mass loss prescription, even with the extra gap torque present. That opposite outcome occurs in the simple mass loss case because the pressure bump is already much further away from the planet.}

%





 
 

\section{Planet migration} \label{sec:migration}

\mklrc{migration can be the last subsection. in the current organization, we go back to fixed orbit runs after migration.}

Planet migration has thus far been the main application of prescribed wind models in 2D planet-disc interactions. One of the most atypical possibilities that has been found is the potential for disc winds to induce outward migration (K20). \cbf{In their work, this outwards migration occurs due to a build-up of excess mass in the co-orbital region in front of the planet and a mass deficit in the co-orbital region behind the planet.} In order to test how disc winds affect planet migration with our model, we explore how the planet migrates both with our default parameters and in cases where outward migration has been found.

\subsection{Fiducial migration} \label{sec:fiducial-migration}

\begin{figure} 
\centering
\includegraphics[width=0.47\textwidth]{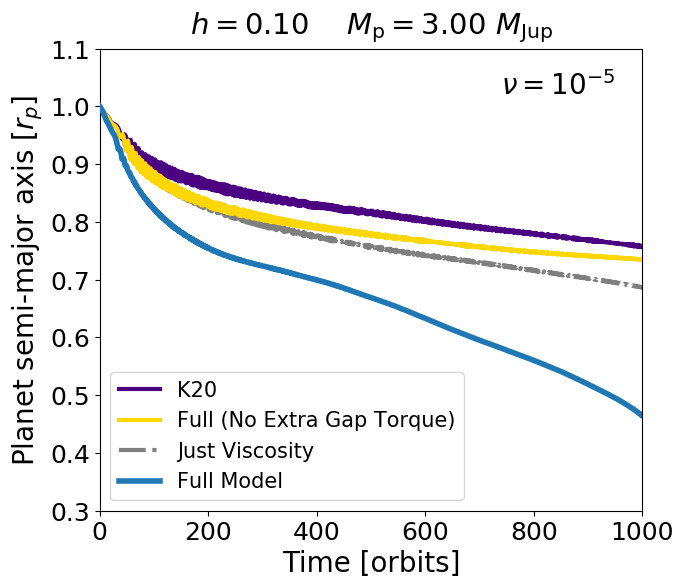} \\
\includegraphics[width=0.47\textwidth]{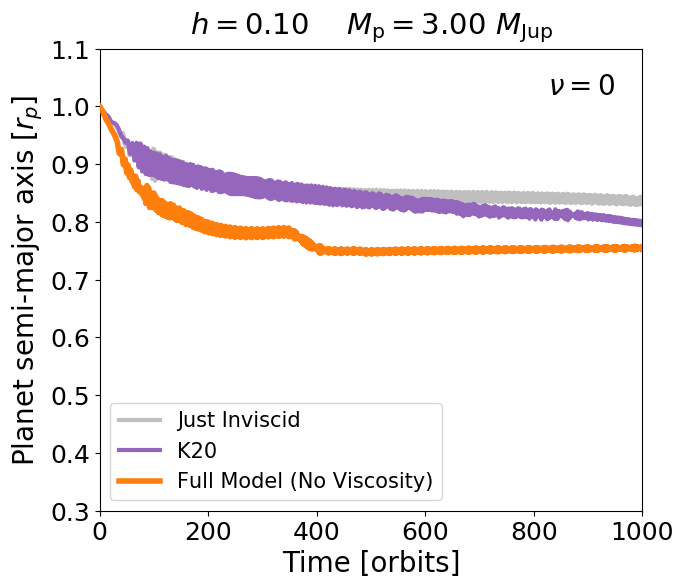}
\caption{Planet migration tracks as a function of time with our fiducial parameters ($h = 0.1$, $\nu = 10^{-5}$, and $M_\mathrm{p} = 3~M_\mathrm{Jup}$). Three main models are featured: [1] this work (\textit{blue, orange, and yellow}), [2] the model from K20 (\textit{purple and dark purple}), and [3] no wind (\textit{grey and dark grey}). The variations leave out the viscosity or extra torque in the gap to illustrate their effects. The planet migrates inwards in all cases, and migrates the fastest with our full model from this work. \cbf{\textit{A migration speed of $1.0~r_\mathrm{p}$ in 1000 orbits corresponds to 2.2 AU / kyr at $r_\mathrm{p} = 20$ AU or 4.4 AU / kyr at $r_\mathrm{p} = 5$ AU.}}}
\label{fig:migration}
\end{figure}

With our fiducial parameters, we find that our prescribed disc wind speeds up migration inwards compared to the K20 model or no wind. \mklrc{inwards or outwards?} We tested two sets of cases, one set with our fiducial viscosity and one with no viscosity. Each set consists of 
\begin{enumerate}
  \item our full prescribed model,
  \item the K20 model
  \item our full prescribed model but with no extra gap torque, and
  \item a control case with just viscosity or no viscosity.
\end{enumerate}
Every setup featured our fiducial $3~M_\mathrm{Jup}$ planet and $ h = 0.1$ disc, while the viscous cases all had \cbf{our fiducial} $\nu = 10^{-5}$. 

As Figure~\ref{fig:migration} shows, the planet migrates inwards in all cases regardless of the model or viscosity, but it migrates the fastest with our full model from this work. We see that the extra gap torque is the dominant factor in making the planet migrate faster, as we can tell from the control case where we removed it. Compared to the general control cases with no wind, the extra gap torque speeds up the migration both with and without viscosity. 

\cbf{In the cases with no viscosity, the migration speed slows down in the later stages. With our model, the migration direction even reverses to a very slow outward pace, a common outcome of planet migration with vortices \citep{lega21, hammer23}.}

\cbf{One caveat is that our model is only based on the outer gap edge and the gap itself, and does not properly take into account the inner gap edge. Future models studying migration could improve on our work by also incorporating the inner gap edge into the development of these models.}

\subsection{Outward migration} \label{sec:outward-migration}

\begin{figure} 
\centering
\includegraphics[width=0.47\textwidth]{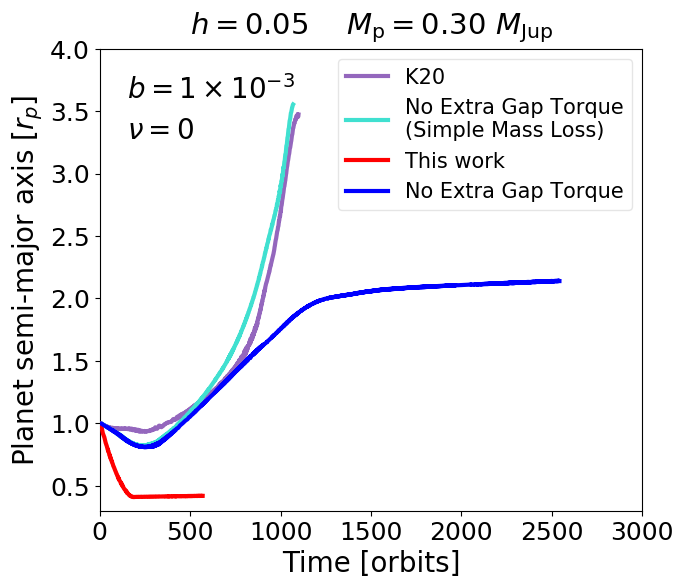} 
\includegraphics[width=0.47\textwidth]{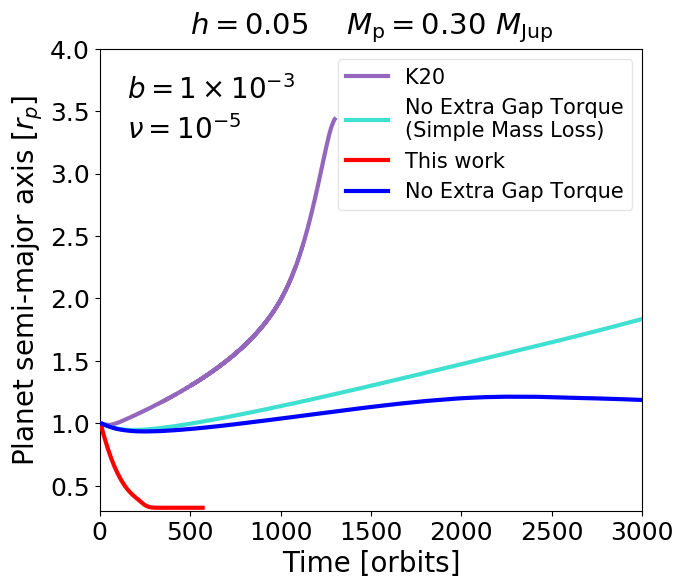}
\caption{Planet migration tracks as a function of time, focusing on a case from K20 with outwards migration ($h = 0.1$, $M_\mathrm{p} = 0.3~M_\mathrm{Jup}$, $b = 10^{-3}$, and $\nu = 0$). With the K20 model and no viscosity (\textit{top panel}), the planet migrates runaway outwards after a few hundred orbits of slow inward migration. With our model, however, the planet migrates inwards rapidly due to the extra gap torque. Even without the extra gap torque, the planet migrates does not begin to migrate outwards until later, and the rate of outward migration is much slower. With viscosity,  (\textit{bottom panel}), the planet still migrates runaway outward with  the K20 model. However, with our model, the viscosity stops the runaway effect and only a very slow outward migration occurs with either mass loss prescription.}
\label{fig:outward}
\end{figure}

To test if the planet can still migrate outwards with our model, we chose parameters from K20 that induced outwards migration. In particular, we lowered the surface density power law to $p = 0.5$ and set the disc aspect ratio to $h = 0.05$. \mklrc{should described the initial surface density profile $r^{-p}$ in the setup}

\mklrc{then just say we focus on inviscid runs if we're not going to show or discuss viscous results}\cbf{Like with our fiducial inwards cases, we tested one set of runs with our fiducial viscosity and another with no viscosity. Each set includes}
\begin{enumerate}
  \item our prescribed model,
  \item the K20 model
  \item our prescribed model but with no extra gap torque, and
  \item our prescribed model but with no extra gap torque and the simpler mass loss prescription from K20.
\end{enumerate}
We focused on Saturn-mass planets with $M_\mathrm{p} = 0.3~M_\mathrm{p}$. With this low of a planet mass, K20 found the planet to migrate outwards with high wind strengths of $b > 4 \times 10^{-4}$. As such, we focus on a relatively high wind strength of $b = 10^{-3}$.

\subsubsection{Inviscid cases} \label{sec:inviscid-outwards}

As Figure~\ref{fig:outward} shows, we observe the planet to migrate outwards using the K20 wind prescription, \cbf{just as they did}. It initially migrates inward slowly before switching to a much faster outward migration. After a little over 1000 orbits, the planet reaches the outer boundary of the domain and then settles there only because that's the edge of the domain. With our model, however, the planet's initial inwards migration is much faster. Whereas the planet in the K20 disc never even reached $r = 0.95~r_\mathrm{p}$, the planet in the disc with our model reaches the inner boundary of the domain in a few hundred orbits. As a result, it does not even get the chance to reverse direction and migrate outward. 

Like with our fiducial parameters, the rapid inward migration with our model is due to the extra gap torque. If we remove the extra gap torque, we see that the planet's inward migration slows down and it only reaches about $r = 0.7~r_\mathrm{p}$ before reversing direction outward. It then proceeds to runaway to the outer boundary like in K20. \cbf{Along the same lines as the runway outwards migration, the reason the planet migrates runaway inwards is due to a mass deficit that develops in front of the planet and a mass excess that develops behinds it, the opposite pattern to what happens in the runaway outwards case. We are unsure why the opposite pattern occurs with extra gap torque.}

Before the outwards runaway occurs, the planet migrates inwards more with our model than with theirs. Once the runaway begins though, the outward migration track ends up very similar, as long as the mass loss profile is the same. We note that the outwards migration tapers off much earlier with our fiducial mass loss profile \cbf{due to faster depletion in the outer disc}; however, we believe that behavior is unrealistic because this mass loss profile is only intended for short-term simulations a few hundred orbits in length. This difference \cbf{in outcomes showcases how it can be better} to stick with the simple mass loss profile from K20 for long-term simulations.

\subsubsection{Viscous cases} \label{sec:viscous-outwards}

\cbf{In the viscous cases, we observe that viscosity can prevent runaway outwards migration, but only with our model and no extra gap torque. Both with our mass loss prescription and the K20 simple mass loss prescription, the planet still migrates outwards, but much more slowly. With simple mass loss, the migration rate slowly increases but never enters the runaway regime. With our prescription, it eventually reverses back inwards as the outer disc depletes, but ultimately the planet's orbit does not change much from where it started. On the other hand, the case with the full K20 model still undergoes outward runaway migration at almost as fast of a rate as the corresponding inviscid case. Similarly, the case with our full model still undergoes runaway inwards migration at almost as fast of a rate as the corresponding inviscid case.}


\cbf{The reason the viscous case with simple mass loss (and the torque profile from our model) no longer migrates runaway outwards is due to the mass excess and deficits never developing. We are unsure why this co-orbital asymmetry never develops, but suggest it may be related to the gap clearing out faster with our model compared to the K20 model. In general, we suspect viscosity can stop runaway migration by clearing out some of the co-orbital material in front of the planet before it can complete its horseshoe orbit.}





\section{Discussion} \label{sec:discussion}

\subsection{Validity of model} \label{sec:validity}

\subsubsection{Basis of the components} \label{sec:basis}

Beyond matching the MHD gap profiles with higher-mass planets, our best-fit model is physically motivated in several different ways. 
First, we use the radial torque profile from the MHD simulations, unlike existing prescribed disc wind models. Second, we have shown that the best fit for the gap profiles uses the exact level of torque inferred from MHD. Third, we incorporate a level of viscosity that is consistent with the MHD simulations, even if it is not possible to infer an exact value for the level of turbulence from the total Maxwell stress in MHD. And fourth, we have extra torque in the gap, as expected from the MHD simulations. These four aspects show our best-fit model is better physically motivated than other prescribed models regardless of how well it can fit the gap profiles.

The only component of our model that is less physically motivated is the mass loss radial dependence on the orbital frequency. We removed this dependence even though it appears the mass loss does scale with the orbital frequency from the MHD simulations. However, we stress that the purpose of removing the scaling isn't to match the background radial dependence. It is just to increase the mass loss near the gap edge. At the gap edge, the mass loss rate is higher in the MHD simulations compared to the rate we would have if we left in the scaling with orbital frequency that decays at larger radii. Removing the scaling was just the simplest way to increase the mass loss rate there. We also note the mass loss background profile does not have a strong impact on the gap profile in the gap or at the outer gap edge in short-term simulations of around a few hundred orbits. For these two reasons, we believe removing the scaling is physically motivated even if the purpose is not to get the right scaling. For long-term simulations, we recommend to use the simpler K20 mass loss prescription instead, as we have done in this study.

\subsubsection{Reference simulations} \label{sec:reference-simulations}

\cbf{Another potential caveat of our model is that it is primarily based on just one simulation from A\&B 2023. We do not believe this is a significant issue because of the similarity of their three main simulations, each with a different planet mass. All three simulations have the same background wind torque power law profile, and roughly the same level of torque enhancement in the gap. The main difference in the torque profiles is the width of the gap region, which we largely account for by scaling it with the Hill radius. The other main difference in the torque is the interaction between the main planetary gap and the secondary inner MHD gap, which we did not account for in any of the cases.}

\cbf{Besides the torque, two other differences between the simulations with different planet masses are the turbulence and mass loss profiles. Likewise, we do not account for these profiles in detail due to their complex non-uniformity in each case. Nonetheless, neither profile has a significant effect on the gap profiles in our study.} \\

\cbf{The other parameter A\&B 2023 varied was the disc's ambipolar Els\"asser number $Am$, which is inversely proportional to the strength of the ambipolar diffusion. They kept a fixed $Am = 3$ in their main study, but ran one additional simulation with $Am = 1$. This stronger level of ambipolar diffusion in the latter case weakened the MRI, lowering the level of turbulence in the disc by about a factor of two. The main difference A\&B 2023 found in this case was the torque enhancement in the gap was cut about in half from $K_\mathrm{gap} \approx 5$ to $\approx 2.5$.} 

\cbf{We ran two additional tests with the parameters used in the $Am = 1$ simulation, leaving $K_\mathrm{gap} = 5$ in one case and lowering it to 2.5 in the other. We can reproduce the gap profile at a delay with either value of $K_\mathrm{gap}$, although our fiducial value of $K_\mathrm{gap} = 5$ actually helps match the gap profile about 100 orbits faster (at 180 orbits vs. 280) than the lower $K_\mathrm{gap} = 2.5$ value tailored to the $Am = 1$ simulation, and at a smaller delay than the match from our fiducial case (at 214 orbits). This discrepancy reinforces that the extra gap torque does not exactly capture why the MHD gap opens faster.}



\subsection{Applicability to simulations} \label{sec:applicability}

We have demonstrated that our model works well for high-mass planets in discs with high aspect ratios with for a specific combination of wind strength and viscosity motivated by the simulations from A\&B 2023. 

Despite that success, we identified three main potential issues with the feasibility of using a prescribed wind model in place of MHD. First, there is a narrow range of wind strengths and viscosities where both parameters affect the gap profile, instead of one dominating over the other. While it has some success in our model, it might not be practical to include both parameters if the ratio between the two were a bit more in favour of the wind or the viscosity. Second, adding in viscosity does not seem to help alter the gap profile in discs with lower aspect ratios in the same way it does at high aspect ratios. Third, with the presence of vortices at the gap edge, the problem of matching the MHD gap profiles reduces to some extent to matching the vortices from MHD. 

Fortunately, none of these issues is necessarily a problem in general.

\begin{itemize}
\item With regard to the difficulties of combining wind strength and viscosity, it may actually be the case that only one of these parameters typically shapes the gap profile. 
\item With regard to the effects of viscosity being different at low aspect ratios, the viscosity was quite high in A\&B 2023 mainly because they allowed MRI. If there were no MRI, viscosity may be less significant. If there were less viscosity, it may not actually affect planetary gap profiles in wind-driven discs or cause any issues, consistent with how our prescribed strong winds can dominate over prescribed weaker viscosities.
\item Lastly, vortices may not be as much of an issue either. Although they dominate the gap profile early on, it is worth noting that with the high viscosity in our study, they disappear fairly quickly in only a few hundred orbits. After they are gone, matching the gap profiles goes back to being the simpler 1D problem. \cbf{In lower-viscosity discs, vortices are expected to survive longer.} Even when they do appear \cbf{though}, they may not be as difficult to match with lower planet masses or lower disc aspect ratios in discs with lower viscosity that are dominated by a wind.
\end{itemize}

Beyond those three issues, another issue to watch out for is the gas not depleting at the Lagrange point like in A\&B 2023. It remains unclear though, whether the gas at the Lagrange point is supposed to deplete in MHD or not.

\cbf{Lastly, we stress the importance of including mass loss into prescribed wind models. A few previous studies of planets in prescribed wind-driven discs only incorporated the torque. We note that if a wind torque is strong enough to govern the evolution of the disc, the corresponding mass loss should likewise have a significant effect on the depletion of the disc for all but the shortest simulations. Despite the importance of the mass loss, we do not know what the mass loss profile should look like in the long-term. For long-term studies of planets in prescribed wind-driven discs, we advocate testing different mass loss profiles to show the robustness of a result.}





\subsection{Applicability to observations} \label{sec:observations}

Disc winds may help to explain unusually wide planetary gaps. As one example, the asymmetric dust trap in Oph IRS 48, the first prominent dust trap ever identified, is unique in several ways \citep{vanDerMarel15b}. One oddity is that the asymmetry is located at 61 AU even though the gas cavity in the inner disc only extends to about 20 AU \citep{vanDerMarel13, bruderer14}. If the asymmetry is indeed a vortex induced by a planet, this would suggest the planet is located about three times further in than the asymmetry, quite a large separation. \mklrc{seems incomplete. how does this relate relate to the present work? A: I hadn't finished this yet.} In a viscous disc, such a large separation could be attained much more easily if there is also a disc wind present (with the simple mass loss prescription) compared to what would happen with no wind, as illustrated in Figure~\ref{fig:long-term}.


\section{Conclusions} \label{sec:conclusions}

In this study, we test the efficacy of prescribed disc wind models in 2D planet-disc hydrodynamic simulations as a means to replicate or expand upon full 3D MHD disc wind simulations that are prohibitively expensive to run over a wide parameter space or for long simulation times. 

By incorporating three main modifications based on the MHD simulations, we were able to reproduce the intermediate-stage gap profiles for planets with sufficiently high mass, but could not replicate the gap profiles for lower-mass planets. The three main changes we made to existing prescribed wind models were incorporating viscosity, adding extra torque in the gap, and flattening the torque radial profile. The viscosity helped match the gap profile at the outer gap edge and along the shoulder by elevating the amplitude of the pressure bump. The extra torque deepened the gap to better match the deeper gaps observed in MHD. The strength of the prescribed wind torque was taken from the MHD simulations, and the level of viscosity is also consistent with the MHD assuming only a small part of the Maxwell stress is turbulent. 

One of the main caveats with our best gap profile fits is that they occur at a noticeable delay compared to the MHD, particularly for the gap depth. We observe that the hydrodynamic gap-opening process lags behind the MHD gap-opening process in general at all times, even at the very beginning. This lagging discrepancy lessens over time as the gap gets deeper. Much of the discrepancy comes directly from gas collecting at the co-orbital Lagrange point behind the planet, something that does not necessarily happen with MHD. We suspect the  purely MHD-driven gap in the inner disc in MHD simulations\mklrc{is it worth reviewing the gap profile obtained by AB23 in sections 2 or 3, where you can highlight the MHD inner gap as a feature? otherwise, it seems like we just mention "the MHD gap" as if the reader should know what we're referring to. A: rephrased here a little bit}, which is not created by the planet and was left out of our fiducial models, also indirectly helps deepen the gap. We find it surprising that the discrepancy occurs so early in the simulations before the gap has even opened, given that the extra torque in the gap mainly occurs due to the gap's reduced density.

With our model showing success for higher-mass planets, we attempted to extend it to longer simulation times and to other parameters, particularly thinner discs. At longer simulation times in viscous discs, we find \cbf{that the outer gap edge can move further away from the planet, the opposite of what has typically been found with existing wind prescriptions. This extra widening of the gap at the outer gap edge occurs mainly due to the extra torque in the gap itself.}
For thinner discs, we note that viscosity is not needed to elevated the pressure bump, as it is already too strong with no viscosity. We caution that this trend also occurs with the low-mass planet gap profiles that we cannot fit. If, however, the shallow gap is the main reason we cannot fit the lower-mass gap profiles, we expect it may still be viable to use our prescribed model for thinner discs as long as the planet is massive enough to open a deep gap.

When we allow the planet to migrate, we observe that the extra torque in the gap drives faster inwards migration proportional to the wind strength. 
Adding in viscosity can also speed up migration, although not by as much as the extra gap torque. The fastest migration rates occur \cbf{when combining} both the extra gap torque and viscosity. In thinner-disc cases where runaway outward migration has previously been found, the extra gap torque provokes the planet to migrate inwards fast enough to avoid ever migrating outwards. \cbf{Even with no extra gap torque, runaway outwards migration can be prevented with viscosity, but we only find that to be the case with our model and not with the K20 model.}


One key to fitting the MHD gap profiles in general is to allow vortices, since these have been observed with MHD. This constraint limits the maximum viscosity that can be used in the model, since higher viscosities can weaken vortices or inhibit vortex formation altogether. It also makes it more difficult to fit the gap profiles in general since they are not truly 1D axisymmetric. Separately, we tested whether our prescribed model allows stronger long-lived vortices to form with high viscosities that would otherwise prevent them in wind-less discs. Even though such stronger vortices had been observed with existing prescribed models, we found that vortices in high-viscosity discs still remained weaker and shorter-lived with our model.



\section*{Acknowledgements}

MH would like to thank Yuhiko Aoyama and Xuening Bai for helpful discussions about their study. MH would like to thank Leonardo Krapp for help with implementing the prescribed models in FARGO3D, and Lina Kimmig for helpful discussions about their study and also helping with the implementation.
\cbf{MH would like to thank the organisers of the \textit{Exoplanets \& Planet Formation Workshop} in Beijing where this work began.} MH would like to thank Kees Dullemond, Richard Nelson, Sijme-Jan Paardekooper, and Paola Pinilla for hosting me to present this work, and the Dustbusters collaboration for organising the \textit{New Heights in Planet Formation} conference where this work was presented. MH also would like to thank Amelia Cordwell, Can Cui, Thomas Rometsch, and Alex Ziampras for helpful discussions. \cbf{We thank the referee for helpful comments and ideas that improved this work.} Computations were performed on the kawas cluster at ASIAA. MH and MKL are supported by the National Science and Technology Council (grants 112-2112-M-001-064-, 113-2112-M-001-036-, 112-2124-M-002-003, 113-2124-M-002-003-, 114-2124-M-002-003-) and an Academia Sinica Career Development Award (AS-CDA110-M06).


\mklrc{Remember to update the acknowledgements.}

\section*{Data Availability}

The data underlying this article will be shared on reasonable request to the corresponding author.



\bibliography{vortex_bibliography}



 \end{document}